\documentclass[journal,12pt,draftclsnofoot,onecolumn]{IEEEtran}
\usepackage{bm} 
\usepackage{multicol}     
\usepackage{multirow}
\usepackage{booktabs}
\usepackage{bigstrut}
\usepackage{hyperref}
\usepackage[table]{xcolor}
\usepackage{amsthm,amsmath,amssymb}
\usepackage{mathrsfs}
\usepackage{color}
\usepackage{balance}
\usepackage{graphicx}
\usepackage{enumerate}
\usepackage{subfigure}
\usepackage{algorithm}
\usepackage{algorithmic}
\usepackage{booktabs}
\usepackage{cite}
\usepackage{amssymb}
\usepackage{amsmath}
\usepackage{textcomp}
\usepackage{amsthm}
\usepackage{authblk}
\usepackage[utf8]{inputenc}
\theoremstyle{remark}
\begin{document}
 
\title{Deep Reinforcement Learning Based Multidimensional Resource Management for Energy Harvesting Cognitive NOMA 
	Communications}

\author{ Zhaoyuan Shi, {\textit{Student Member, IEEE}},  Xianzhong Xie, {\textit{Member, IEEE}},  Huabing Lu, {\textit{Student Member, IEEE}}, Helin Yang, {\textit{Member, IEEE}},  Jun Cai, {\textit{Senior Member, IEEE}, and Zhiguo Ding,  {\textit{Fellow, IEEE}}
	}
	
\thanks{ 
	This work was supported by the National Nature Science Foundation of China under Grant No. 61502067, the Key Research Project of Chongqing Education Commission under Grant No.  KJZD-K201800603, the Key Project of Science and Technology Research of Chongqing Education Commission under Grant No. KJZD-M201900602, the Chongqing Nature Science Foundation under Grant No. CSTC2018jcyjAX0432 and CSTC2016jcyjA0455, Key Project on Anhui Provincial Natural Science Study by Colleges and Universities under Grant No KJ2020A0514, KJ2020A0513, KJ2020A0497, and KJ2019A0554, and the Doctoral Candidate Innovative Talent Project of Chongqing University of Posts and Telecommunications under Grant No. BYJS201912.
		
	Z. Shi is with School of Computer Science and Technology,   Chongqing University of Posts and Telecommunications, Chongqing, China and Key Laboratory of Intelligent Perception and Computing of Anhui Province, Anqing Normal University, Anqing,  China (email:shizy@stu.cqupt.edu.cn).
	
	X. Xie and  H. Lu are with Chongqing Key Lab of Computer networks and Communication Technology.Chongqing University of Posts and Telecommunications, Chongqing, China(email:xiexzh@cqupt.edu.cn, ai4b@163.com).
	
	H. Yang is with the School of Electrical and Electronic Engineering,
	Nanyang Technological University, 50 Nanyang Avenue, 639798, Singapore,
	(e-mail: hyang013@e.ntu.edu.sg).

	J. Cai is with the Network Intelligence and Innovation Lab (NI$^2$L), Department of Electrical and Computer Engineering, Concordia University, Montreal, QC H3G 1M8, Canada (e-mail: jun.cai@concordia.ca).
	
	Z. Ding is with the Department of Electrical Engineering, Princeton University, Princeton, NJ 08544, USA. And the School of Electrical and Electronic Engineering, the University of Manchester, Manchester, UK (email:
	zhiguo.ding@manchester.ac.uk).

}
}

\maketitle

\begin{abstract}
The combination of energy harvesting (EH), cognitive radio (CR), and non-orthogonal multiple access (NOMA) is a promising solution to improve energy efficiency and spectral efficiency of the upcoming beyond fifth generation network (B5G), especially for support the  wireless sensor communications in Internet of things (IoT) system.  However, how to  realize intelligent frequency, time, and energy resource allocation to support better performances is an  important problem to be solved. In this paper, we study joint spectrum, energy, and time resource management for the EH-CR-NOMA IoT systems. Our goal is to minimize the number of data  packets losses for all secondary sensing users (SSU), while satisfying the constraints on the maximum charging battery capacity, maximum transmitting power, maximum buffer capacity, and minimum data rate of primary users (PU) and SSUs. Due to the non-convexity of this optimization problem and the stochastic nature of the wireless environment, we propose a distributed multidimensional resource management algorithm based on deep reinforcement learning (DRL).   Considering the continuity of the resources to be managed, the deep deterministic policy gradient (DDPG) algorithm is adopted, based on which each agent (SSU) can  manage its own  multidimensional  resources without collaboration.  In addition, a simplified but practical action adjuster (AA) is introduced for improving the training efficiency and battery performance protection. 
The provided results show that the convergence speed of the proposed algorithm is about 4 times faster than that of  DDPG, and the average number of packet losses (ANPL) is about 8 times lower than that of the greedy algorithm.

\end{abstract}
\vspace{-10pt} 

\IEEEpeerreviewmaketitle
\section{Introduction}

  \renewcommand{\baselinestretch}{0.75} 
According to Cisco report (2018-2023), there will be 3.6 networked devices per capita by 2023, up from 2.4  in 2018. There will be 29.3 billion networked devices by 2023, up from 18.4 billion in 2018. The rapid development of communication technologies has realized the era of interconnection of everything (IoE) \cite{cisco}. However, the accompanying shortage of spectrum resources and the dramatic increase in energy consumption have also attracted widespread attention. 

Cognitive radio (CR) and non-orthogonal multiple access (NOMA) have gained tremendous research interests as promising paradigms due to their outstanding performance in improving spectrum efficiency and providing massive connectivity \cite{crnoma}. In the CR network, the unlicensed secondary users (SUs) are allowed to opportunistically access the channels of the licensed primary users (PUs) if collisions or harmful interference can be effectively avoided \cite{CR}. As a promising multiple access technology in upcoming the fifth-generation (5G) communication systems and beyond application, NOMA allows multiple users to share the same time and spectrum resources with different power levels, and separates the multi-user signals by applying  successive interference cancellation (SIC) at the receivers \cite{NOMA}.  In recent years, several works have been carried out to investigate the CR-NOMA system \cite{ADD_R6,ADD_R7}. In \cite{ADD_R6}, Xiang \textit{et.al} designed a novel secure transmission scheme for hybrid automatic repeat request (HARQ) assisted CR-NOMA networks. In \cite{ADD_R7},  LV et al. detailed the advantages of combining CR and NOMA, i.e., improving spectral efficiency, providing large-scale connectivity, low latency, better fairness, etc., then presented three different CR-NOMA networks, and finally pointed out some challenges in CR-NOMA networks, such as interference cancellation, non-perfect CSI, etc. 

Energy harvesting (EH) \cite{EH} is another key technology for the next generation  networks, which can provide controllable energy supply and prolong the lifetime of energy-constrained networks. A variety of green energy sources such as light, heat, wind, and radio frequency (RF) can be utilized for EH. In recent years, with the development of the Internet of Things (IoT) and wireless sensor networks (WSN), harvesting energy from ambient RF signals has attracted a lot of attention because it enables low-power communication networks to be energy self-sustainable \cite{EH2}. 

The convergence of EH, NOMA, and CR, known as EH-CR-NOMA, can effectively address some of the challenges of 5G, including  spectrum resource utilization improvement, energy consumption reduction, and massive connectivity while maintaining reliable communication between PUs and SUs. 
In recent years, different combinations of CR, EH, and NOMA technologies have been extensively studied. In \cite{EHCR,EHCR1,EHCR2}, EH and CR were combined and investigated. In \cite{EHCR}, a novel channel selection mechanism was proposed for EH-CR networks, where each SU harvests  RF energy from the active PUs and transmits data in other selected idle channels. Zhou \textit{et al.} 
developed a 3D matching algorithm for EH-CR machine-to-machine networks that maximizes the energy efficiency of the system \cite{EHCR1}. An Lyapunov optimization-based resource allocation algorithm was introduced for EH-CR WSN in \cite{EHCR2}. Simultaneous wireless information and power transfer (SWIPT) with NOMA was studied in \cite{EHNOMA,EHNOMA1}, where the SWIPT is a special case of EH. In \cite{EHNOMA}, a novel user pairing and power allocation algorithm was proposed for the SWIPT-NOMA system, which can maximize  spectral and energy efficiency. 

In EH-CR-NOMA systems and other EH systems, resource management has always been a hot research topic from the perspective of green communication.  
Specifically, it is very important to study how to efficiently utilize the harvested energy and make a good trade-off between EH and data transmission. Tang \textit{et al.} developed a dual-layer energy efficiency optimization algorithm for the SWIPT-NOMA system in \cite{EHNOMA1}, in which the Dinkelbach method was used to optimize the power allocation and control time switching allocation, and considering equal time switching factors in all terminals. In \cite{EHNOMACR} Li \textit{et al.} studied a resource allocation scheme in the EH-CR-NOMA scenario, in which the SUs collect the RF energy from the PUs, and the harvested energy is greedily used up for the forwarding of the PUs information and the signal transmission of the SUs. The authors developed a two-level bisection search algorithm to select the time portion for energy harvesting and the transmit power coefficients in NOMA transmission. However, this algorithm is only applicable to CR systems with a single PU-SU pair. In \cite{EHNOMACR1},  the authors used the Dinkelbach method to maximize the energy efficiency of an EH-CR-NOMA system, where in the overlay CR mode the SU needs to detect the spectrum occupancy of the PU. However, the \cite{EHNOMA1} studied only one scenario in which the PU will not occupy the channel all the time and assumed that the spectrum detection is perfectly accurate or just perfect, which is not realistic in the actual CR system. A Dinkelbach algorithm was applied to find the optimal resource management policy in EH-WSN \cite{EHNOMACR3}, which divides the time slot into two parts, the first part for EH, and the second part for sensor data transmission. In \cite{EHoff}, an energy-aware traffic offloading scheme was proposed to minimize the power consumption while satisfying the system quality of service (QoS) requirement for an EH heterogeneous cellular network (HCN).

Most of the above assume that the system channel information and the harvested energy are perfectly known and controllable. However, such an assumption is often unrealistic in EH wireless communication systems due to the stochastic property of the wireless environments and the harvested energy.  In addition, considering the security factors and the load of information interaction, it is impractical to realize the sharing of information among users in the time-varying EH system.   Thanks to the recent advances in artificial intelligence, the paradigm of deep reinforcement learning (DRL) has provided a promising approach to address this issue. Due to the powerful learning capability in dynamic unknown environments, DRL is widely adopted to learn the optimal decision policy in wireless communications \cite{DRL_SUV}, including resource management \cite{DRL_power,DRLres3,ziji_iot}, dynamic spectrum access \cite{DRLres1,DRLres2}, and data offloading \cite{offload,offload1}. 

With the development of EH technology, DRL has also been introduced into the EH wireless communication systems to  the optimization resource management \cite{EHDRL,EHDRL1,new,EHDRL2,EHDRL3,duibi_lett,duibi_ding}. In particular, in \cite{EHDRL}, a reinforcement learning (RL)-based power allocation algorithm was studied for maximizing the long-term sum rate of an underwater full-duplex EH relay network consisting of a source, a destination, and a relay. In \cite{EHDRL1}, Chu \textit{et al.} proposed a DRL-based resource scheduling algorithm for EH IoT systems, where the unique base station (BS) in the system acts as the central agent to schedule the dynamic access and transmission power of all EH users. In \cite{new}, Zhao  \textit{et al.} integrated energy harvesting with wireless local area networks (WLANs) and proposed a DRL-based access control algorithm. A novel energy management algorithm based on the  deep deterministic policy gradient (DDPG) was investigated in \cite{EHDRL2}, which considered both the EH point-to-point network and the EH one-way relay network. In \cite{EHDRL3}, Min \textit{et al.} proposed an RL-based offloading algorithm for 
choosing the edge device and the offloading rate in an EH IoT system. The algorithm allows EH device to achieve intelligent offloading without knowledge of MEC, energy consumption, and computational latency. 

In the EH system, the problem of joint time and transmit power optimization based on DRL was studied in  \cite{duibi_lett, duibi_ding}. Due to the outstanding performance advantage in the continuous action space problem, the DDPG \cite{ddpg} algorithm was used in \cite{duibi_lett,duibi_ding} for continuous time and transmit power decisions. In \cite{duibi_lett},  Li \textit{et al.}  studied an EH point-to-point network, in which rechargeable batteries were installed in the transmitter. Based on the  novel DDPG algorithm, the transmitter sends data with an appropriate transmit power in the period $\tau_t$, and harvests energy in the remaining time slot $T-\tau_t$. To the best of our knowledge, the joint DRL-based time and power scheduling problem was first studied in EH-CR-NOMA systems in \cite{duibi_ding}. Ding \textit{et al.} proposed a novel DDPG-based resource allocation algorithm to maximize the long-term throughput of SUs, where the only SU, as an agent, learns the allocation strategy of transmit power and energy collection duration through   exploration and training.

The above DRL-based algorithms can achieve good system performance in EH wireless communication systems without knowing the prior information of the system in advance. However, few papers have studied the joint scheduling of frequency resources, energy resources, and time resources in EH systems. Besides, these algorithms are only applicable to EH systems with a simpler communication model, where there is a single EH user \cite{EHDRL,EHDRL2,EHDRL3,duibi_lett,duibi_ding}.  Although \cite{EHDRL1} considers the case with multiple EH users, and the BS in this system acts as an agent for centralized scheduling of all EH users,  the process requires additional information interaction overhead for state acquisition and decision execution.  Also, except 
\cite{duibi_ding}, there is little work on joint resource scheduling based on the DRL algorithm for EH-CR-NOMA systems.

Therefore, we have integrated CR, NOMA and EH technologies. They can facilitate the self-sustainability of energy-limited sensor users, as they can be able to share the spectrum and harvest the energy from the ambient RF sources. Because CR allows them to share spectrum resources with licensed PUs, and EH allows them to recharge from the surrounding RF environment. NOMA technology enables simultaneous access of multiple sensor users.

Motivated by the above,  and considering the actual communication system scenario, in this paper,  we investigate a more practical green communication system for  sensor users in IoT, which contains multiple EH secondary sensor users (SSUs) that collect RF energy from the PUs for their own data transmission. In this system, low-cost and energy-constrained unlicensed SSUs share the spectrum resources of licensed PUs through CR technology and use NOMA  to provide more connections to more SSUs, in addition to which EH is used to ensure the energy sustainability  of SSUs.  In particular, a joint management scheme of time resource, energy resource, and spectrum resource based on a distributed DDPG algorithm is proposed for the system. The main contributions of our work are summarized as follows:
\begin{itemize}
\item To achieve timely data communication for multiple users with limited communication resources, a more practical EH-CR-NOMA IoT framework is developed, which contains multiple PUs and multiple EH  SSUs, and each SSU needs to transmit the data packets sensed from the IoT environment to the BS in time, due to the limited space of the configured buffer. Each time slot is dynamically divided into two parts, in the former part, each SSU transmits the collected data to the BS  with the energy in the rechargeable battery,  and in the latter part the SSU harvests and stores RF energy from the PUs.

\item We propose a distributed DDPG-based multidimensional resource management algorithm to achieve the optimization goal of minimizing the average number of  packets losses (ANPL)  in the system.  Unlike \cite{DRL_power, DRLres2}, the state information in the proposed algorithm does not require any collaboration between agents, nor does it need to use long-short-term
memory (LSTM)  technology for state observation as in \cite{DRLres1}. The state in the proposed algorithm is based only on locally observable information. Besides, a reward function that takes into account the SSU throughput, the number of packets losses, and the QoS of the PUs is designed to drive the joint management of time resources (communication duration and energy harvesting duration allocation), energy resources (transmit power control), and frequency resources (dynamic spectrum access) for each SSU.

\item Considering the limited computational capacity of each SSU, we adopt a centralized training distributed execution method for the proposed  algorithm, which can simplify the implementation of the algorithm and improve its stability. In addition, to further accelerate the convergence speed and improve the performance of the proposed DDPG algorithm, we introduce the action adjuster (AA) and named the improved algorithm as AADDPG.  Simulation results well demonstrate the robustness and performance advantage of the proposed AADDPG algorithm in terms of the average reward,  number of  packets losses  sum rate of SSUs,  energy efficiency, and the algorithm convergence speed.


\end{itemize}

The remainder part of this paper is organized as follows. Section \uppercase\expandafter{\romannumeral2} describes the system model. In Section \uppercase\expandafter{\romannumeral3}, the proposed DRL-based framework is demonstrated. Section \uppercase\expandafter{\romannumeral4} presents the details of proposed AADDPG algorithm. The simulation results and concluding remarks are finally presented in Section \uppercase\expandafter{\romannumeral5} and \uppercase\expandafter{\romannumeral6}, respectively. 

\textit{Notations}:  For ease of understanding, we define the symbols with subscripts $p$ and $s$ to belong to PUs and SSUs, respectively. We use $\lfloor \cdot\rfloor$ to denote the downward rounding operation,  and use lowercase bold fonts to represent vectors.
\section{System Model}

\begin{figure}[t]
	\centering
	\includegraphics[width=5.5in]{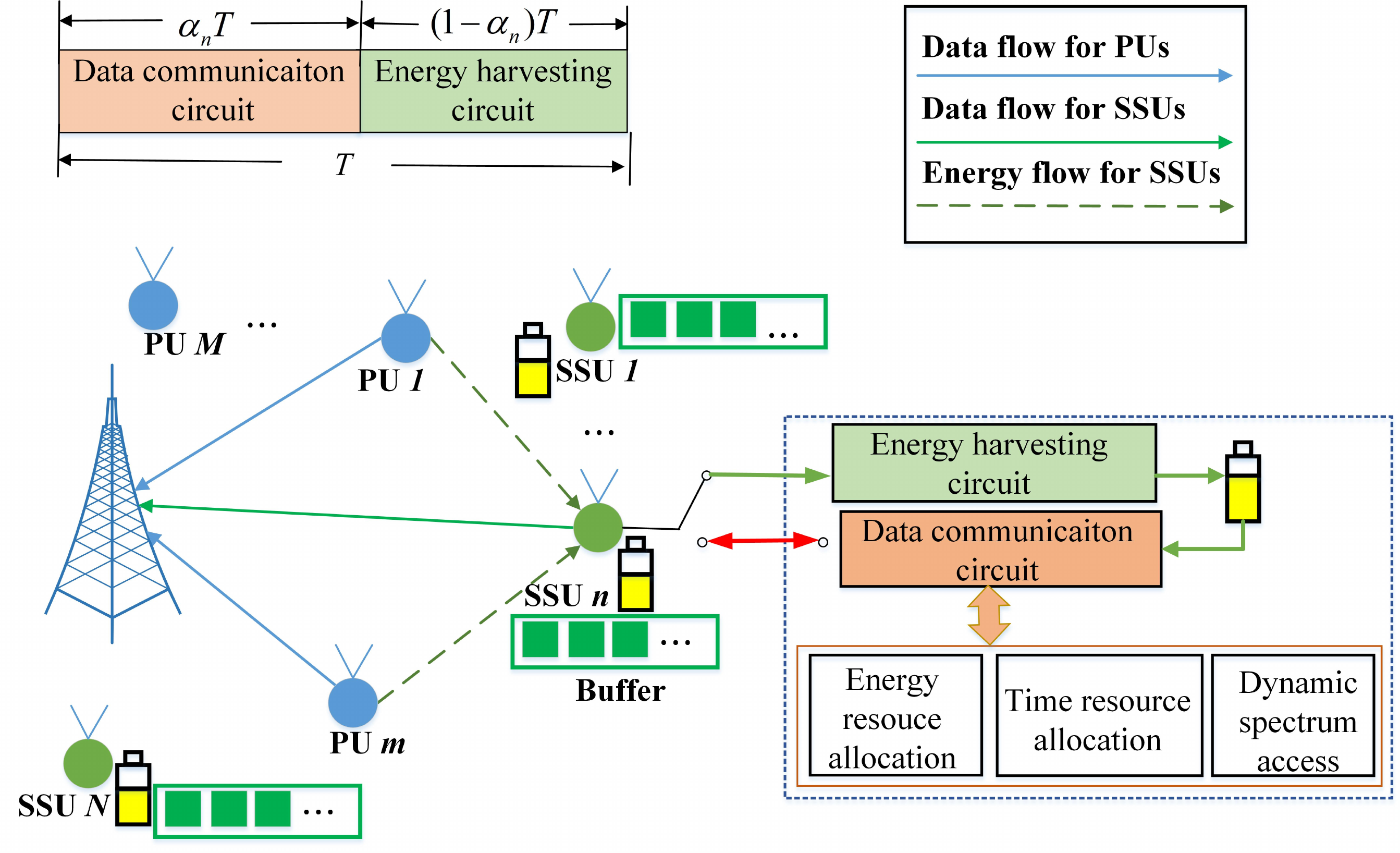}
	\caption{The communication model of EH-CR-NOMA IoT system }
	\centering
	\label{Fig1}	
\end{figure}

An uplink EH-CR-NOMA IoT system is considered as shown in Fig. \ref{Fig1}, which consists of one  BS, 
a set $\mathcal{M}=\{1,2,\ldots,M\}$ of PUs, and a set $\mathcal{N}=\{1,2,\ldots,N\}$ of EH SSUs. Each SSU equips a separate environmental data acquisition unit, and the collected data is temporarily stored in a buffer. Due to the limited capacity of the buffer, the SSU will use the harvested energy to transmit the collected data to the BS as soon as possible. Note that the data collection process is independent of energy harvesting and data transmission.   Each node in the network is assumed to be equipped with a single antenna as
\cite{EHNOMA1,EH2,new,duibi_lett,duibi_ding}. 	
We assume that each PU decides its transmission based on two-state Discrete-Time Markov Chain (DTMC) model, where the transfer probabilities are $P_1$ and $P_2$. \footnote{Assuming that the current active state of the PU is busy (communicating) then at the next time slot, the PU will switch to the idle state with probability $P_1$ and remain busy with probability $1-P_1$. Conversely, if the PU is in the idle state currently, then it will change to the busy state with probability $P_2$  and remain in the idle state with probability $1 - P_2$ at the next moment.}   For SSUs, underlay access mode is used, i.e., the accessed SSUs will share the same channel in the system with the PUs, provided that no harmful interference is caused to the PUs.  Note that both PUs and SSUs are randomly distributed in a certain range around the BS. However, to guarantee the performance of PUs, we set the SSUs much farther away from the BS than the PU.

 The channel in the system contains small-scale Rayleigh fading and large-scale path loss. Specifically, the channel  between the SSU $n$ (PU $m$) and the BS is denoted as $h_s^n=\sqrt{\beta_s^n} \widetilde {h_s^n}$ ($h_p^m=\sqrt{\beta_p^m} \widetilde {h_p^m}$), where  $\widetilde{h}_s^{n}\sim \mathcal{CN} (0,1)$ and $\beta_s^{n}=\left(\dfrac{\lambda^\ast}{4\pi d_s^n}\right)^2$ $\left(\beta_p^{m}=\left(\dfrac{\lambda^\ast}{4\pi d_p^m}\right)^2\right)$ account for the Rayleigh fading and path loss coefficient, respectively.  $\lambda^\ast$ is the signal wavelength, and $d_s^n$ ($d_p^m$) presents the distance between the $n$th SSU ($m$th PU) and BS. Similarly, we define the channel between the PU $m$ and the SSU $n$ as $g_{mn}=\sqrt{\beta_{mn}}\widetilde{g_{mn}}$. We assume a quasi-static synchronized time slotted system with slot duration $T$: each channel keeps constant within each time slot, and change independently between time slots. For simplicity, a single frequency band  of $BW$ Hz is considered  in this paper.

A transmit-then-harvest protocol is employed as \cite{duibi_ding}, under which each time slot $t$ consists of two phases: data transfer and EH. In particular, the $n$th SSU transmits the  data  packets to BS in the first $\alpha_n(t)T$, and performs battery charging by harvesting the ambient RF energy from PUs in the remaining $(1-\alpha_n(t))T$,  where  $\alpha_n(t)$ is a time-sharing parameter with $0\leq\alpha_n(t)\leq 1$.  All transmitting users access the same frequency resource block via the NOMA protocol. Let binary indicator $U_p^m$ and $U_s^n$ denote the active state of the $m$th PU and the access state of $n$th SSU, respectively. Then, the signal received at the BS can be calculated as 
\begin{equation}
y(t)=\sum_{n=1}^{N}U_s^n(t)\sqrt{p_s^n(t)}h_s^n(t)x_s^n(t)+\sum_{m=1}^{M}U_p^m(t)\sqrt{p_p(t)}h_p^m(t)x_p^m(t)+z,
\end{equation}
where $p_p$ is the transmit power of each PU, $z$ denotes the zero-mean additive white Gaussian noise with variance $\sigma^2$. $x_s^n$ and $x_p^m$ are the transmit message of the $n$th SSU and the $m$th PU, respectively, with $\mathbb{E}\{|x_s^n|^2\}=\mathbb{E}\{|x_p^m|^2\}=1$. The transmit power of SSU satisfies $p_s^n(t)\in[0,p_s^{max}]$, where $p_s^{max}$ is the maximum transmit power of all SSUs.
\vspace{-2pt}
\subsection{Energy Harvesting Protocols}

 In the uplink EH-CR-NOMA system studied in this paper, SSUs use the time duration $(1-\alpha_n(t))T$ for RF energy harvesting in each time slot. For the green communication perspective, we do not provide an additional source of RF energy supply for SSUs, such as BS. Instead, SSUs collect the RF energy generated by PUs in the environment for communication. Note that the energy harvesting operation of SSUs does not cause additional energy consumption of PUs \cite{EH_sey}.  Without loss of generality,  we define $E_n(t)$ as the remaining energy in $n$th SSU's battery at the beginning of time slot $t$.   \footnote{For simplicity, we ignore the energy consumed by circuitry and signal processing, but our algorithm can be easily extended to include these energy-consuming scenarios.} Based on energy harvesting protocols, for the $n$th SSU,  $\alpha_n(t)Tp_s^n(t)U_s^n(t)$ energy will be consumed for the data transmission,  and in EH phase, $U_s^n(t)(1-\alpha_n(t))T\eta\sum_{m=1}^{M}U_p^m(t)\mid g_{mn(t)}\mid^2p_p(t)$ energy will be harvested, where $\eta$ stands for the EH efficiency coefficient.  Thus, $E_n$ can be evolved as
\begin{equation}
\begin{aligned}
&E_{n}(t+1)=\\ &\text{min}\{U_s^n(t)(1-\alpha_n(t))T\eta\sum_{m=1}^{M}U_p^m(t)p_p(t)\mid g_{mn}(t)\mid^2+E_n(t)-\alpha_n(t)Tp_s^n(t)U_s^n(t), E_{max} \},
\end{aligned}
\label{en}
\end{equation}
where $E_{max}$  denotes the capacity limit of the SSU's battery. Based on the communication protocol of SSUs, i.e., the SSUs transmit data first and harvest the energy later, we can limit SSU's  transmit energy as follows
\begin{equation}
0\leqslant\alpha_n(t)Tp_s^n(t)U_s^n(t)\leqslant E_n(t).
\label{power}
\end{equation}


\subsection{Transmission of Collecting Data Packets}
In this paper, each SSU contains a separate data acquisition device for environmental data collection.  The collected data is temporarily stored in a buffer and then transmitted to the BS by using the harvested energy. Denote $c_n(t)$ as the number of data packets collected by the $n$th SSU during time slot $t$, which obeys an 
independent Poisson distribution process with an identical rate $\lambda$.  We define $l_n(t)$ as the number of packets successfully transmitted to the BS,  which can be expressed as follows:
\begin{equation}
\begin{aligned}
l_n(t)=\min\left[\lfloor \dfrac{BW\alpha_n(t)T\text{log}_2\left(1+{\rm SINR}_n(t)\right)}{C} \rfloor, B_n(t)\right],
\end{aligned}
\end{equation} 
where ${\rm SINR}_n$ stands for the signal-to-interference-plus-noise (SINR) per unit bandwidth of the $n$th SSU. $C$ is the number of bits contained in each data packet, and $B_n(t)$ denotes the queue length of buffer of the $n$th SSU at the beginning of time slot $t$. Therefore, $B_{n+1}(t)$ can be expressed as follows:
\begin{equation}
B_{n+1}(t)=\text{min}\{(B_n(t)-l_n(t)+c_n(t)),B_{max}\}, \quad n\in \mathcal{N},
\label{bn}
\end{equation}
where $B_{max}$ is capacity limit of each SSU's  buffer.  Obviously, buffer overflow occurs when the collected data packets cannot be transmitted in time. This results in packet loss and the number of lost packets is 
\begin{equation}
L_{n}(t)=\left(B_n(t)-l_n(t)+c_n(t)\right)-B_{max}, \quad n\in \mathcal{N}.
\end{equation}

\vspace{-2pt}
\subsection{Successive Interference Cancellation  (SIC) Decoding}
To decode the received data in the EH-CR-NOMA system, SIC is carried out at BS. Specifically, the BS determines decoding order based on the signal strength, which depends on the transmitted power and channel gain of users.  Note that the channel state information (CSI) is required for the SIC decoding process \cite{NOMA,sic_fail_next}, which can be obtained by pilot signals \cite{DP}.  The user with the strongest signal strength will be decoded first. Based on the principle of SIC,  each decoded user's signal will be regenerated and then subtracted from the remained signal. The signals of those users who failed to be decoded and also those who have not been decoded will be both regarded as the interference \cite{sic_fail_next}. For the $i$th decoded user with a signal strength of $g_i=U_s^i(t)\mid h_s^i(t)\mid^2p_s^i(t), i\in\mathcal{N}$ or $g_i=U_p^i(t)\mid h_p^i(t)\mid^2p_p(t),  i\in\mathcal{M}$, it will be subject to interference from SSUs that have not yet decoded or have failed to decode as
\begin{equation}
\begin{aligned}
I_s^i(t)=\sum_{j\in\mathcal{N},j\neq i}U_s^j(t)\left[1-\beta_s^{ji}(t)d_s^j(t)\right]|h_s^j(t)|^2p_s^j(t),
\end{aligned}
\end{equation}
where, $\beta_s^{ji}$ ($\beta_p^{ji}$) is a binary indicator with  $\beta_s^{ji}=1$ ($\beta_p^{ji}=1$) if the signal strength of the $j$th SSU (PU) is greater than  the currently decoded user $i$, i.e., $|h_s^j(t)|^2p_s^j(t)>g_i$ ($|h_p^j(t)|^2p_p(t)>g_i$), and $\beta_s^{ji}=0$ ($\beta_p^{ji}=0$) otherwise. Besides, we use $d_s^j(t)=1$ ($d_p^j(t)=1$) to indicate that the $j$th SSU (PU) has been successfully decoded, and $d_s^j(t)=0$ ($d_p^j(t)=0$) to indicate that it has failed to decode or has not been decoded. Similarly, we can describe the interference from the PUs as follows
\begin{equation}
I_p^i(t)=\sum_{k\in\mathcal{M},k\neq i}U_p^k(t)\left[1-\beta_p^{ki}(t)d_p^k(t)\right]|h_p^k(t)|^2p_p(t).
\end{equation}
Thus, the SINR of user $i$ can be written as 
\begin{equation}
{\rm SINR}_i(t)=\dfrac{g_i}{I_s^i+I_p^i+\sigma^2} .
\end{equation}
Then the achieved data rate of the $i$th user can be expressed as:
\begin{equation}
\begin{aligned}
&R_p^i(t)=\text{log}_2(1+{\rm SINR}_i(t)), \quad &\forall i\in \mathcal{M}, 
\end{aligned}
\end{equation}
 To meet the QoS requirements of PUs, the rate of the PUs must  satisfy 
 $R_p^i(t)>R^1$, where $R^1$ is the rate threshold of each PU.  In addition, To decode SSUs successfully, the data rate of SSU $n$ should satisfy $R_s^n>R^0$, where $R^0$ denotes the  threshold of all SSUs. 
\subsection{Optimization Problem Formulation}
 In this paper,   SSUs are focused on and they tend to have the following characteristics: smaller battery capacity, limited storage space, and lower computational power. Therefore, we need to design the dynamical resource management algorithm to achieve the continuity of energy and timely successfully access for SSUs. 
Specifically, we focus on the problem of minimizing the number of loss packets of all SSUs for the EH-CR-NOMA IoT system,  by optimizing the access indicator, $U_s^n$ (dynamic spectrum access), the time-sharing factor, $\alpha_n$ (time resource allocation), and the transmit power $p_s^n$ (energy resource management). The optimization problem can be formulated as follows
\begin{equation}
\nonumber
\begin{aligned}
\textbf{(P1):}\quad&\min_{U_s^n,p_s^n,\alpha_n} \sum_{n=1}^{N}L_n(t);\\
\text{S.t. : }&U_s^n(t),U_p^m(t)\in \{0,1\}, m\in \mathcal{M}, n\in \mathcal{N} ; & (P_a) \\
&p_s^n(t)\in\left[0, \text{min}\left\{p_s^{max},\dfrac{E_n(t)}{\alpha_n(t)T}\right\}\right], (\ref{power});& (P_b)\\
&0\leq\alpha_n(t)\leq 1,\quad 0\leq B_n(t)\leq B_{max}&(P_c),\\
&R_p^m(t)\geq R^1,\quad  R_s^n(t)\geq R^0,  m\in \mathcal{M}, n\in \mathcal{N}; &(P_d)\\
&(\ref{en}),(\ref{bn});&(P_e)
\end{aligned}
\label{opt}
\end{equation}
where $(P_b)$ assures that  the transmit power $p_s^n$ is no greater than both the maximum transmitting power $p_{max}$  and the maximum power that can be supplied by the battery. $(P_e)$ gives the principles for the evolution of energy value $E_n$ and  buffer length $B_n$. $(P_d)$ presents the QoS requirements of PUs and SSUs. To achieve the optimization goal, when the remaining battery power of a SSU is low, its transmit power may be set to 0; such that it can harvest energy throughout the time slot; When the buffer of a SSU is almost full, it will strive to use more transmit power to achieve successful communication. That is, to achieve the optimization goal, each SSU needs to learn to make intelligent decisions about the $U_s^n$, $p_s^n$, and $\alpha_n$ based on the local information it observes, such as remaining energy, buffer length, channel quality, etc., which can be realized by the DRL algorithm. This will be discussed in the following two sections.

\section{Continuous  Drl Formulation}

Considering the non-convexity and complexity of the optimization problem in ($\textbf{P1}$)   which cannot be solved by traditional convex optimization algorithms, together with the fact that the environmental state transfer probability is expected to be unknown, we develop a DRL-based strategy to implement the joint management of time, energy and frequency resources for the  EH-CR-NOMA IoT system.

\vspace{-2pt}
\subsection{Agent and State Space}
Unlike the literature \cite{EHDRL,EHDRL2,EHDRL3,duibi_lett,duibi_ding}, the communication model we study contains multiple EH users. Without a lot of information exchange, it is difficult to achieve  multidimensional resource management of all SSUs with a single  central agent in the developed EH-CR-NOMA IoT system. Considering the costly information exchange load  in the centralized resource scheduling algorithm, a multi-agent distributed DRL algorithm is designed in this paper. Specifically,  as an independent agent, each SSU makes its own resource management decisions.

In each time slot $t$, the agent $n,n\in\mathcal{N}$, observes the environment and obtains the state vector $\textbf{s}_n(t)$, according to which an action $\textbf{a}_n(t)$ will be executed and the corresponding reward $R_n(t)$ will be obtained. However, unlike the centralized learning with a single agent, the state transitions and the reward obtained in the multi-agent DRL algorithm depend on the joint actions of all the agents, which leads to potential instability in the learning process. To solve this problem, it is often necessary to have more information exchange among all agents as in 
\cite{DRL_power, DRLres2}, which inevitably imposes an additional burden, especially  when the number of agents is large. Alternatively the LSTM method could be adopted to obtain the state information of accumulated time period as in \cite{DRLres1}. In this work, we take another centralized training, and distributed execution DRL approach to simplify algorithm implementation and improve stability, which will be discussed in Section \uppercase\expandafter{\romannumeral4}.

In the proposed distributed DRL-based resource management algorithm,  the state information is based only on local observation, that is, each SSU does not need to know the information of other SSUs, such as channel status information (CSI), battery power information, buffer length, transmission power, etc., which greatly reduces the system overhead. Specifically, the  environment state vector of agent $n$ is defined as 
\begin{equation}
\textbf{s}_n(t)=\left[h_s^n(t),\mathbf{U}_p(t),\mathbf{G}_n(t), E_n(t), B_n(t)\right],
\label{state}
\end{equation}
where $\mathbf{U}_p(t)=[U_p^1(t),U_p^2(t),\ldots,U_p^M(t)]$ represents the communication status of all PUs at time $t$, $\mathbf{G}_n(t)=[g_{1n}(t),g_{2n},\ldots,g_{Mn}(t)]$, denotes the channel information between all PUs and SSU $n$.
We assume that the communication status of all PUs can be immediately detected by each SSU \footnote{It can be achieved with the already sophisticated signal detection technology.}.
Obviously,  the channel information $h_s^n(t)$, $\mathbf{G}_n(t)$,  battery information $E_n(t)$, and buffer information $B_n(t)$ can be locally observed.
Note that this comes without requiring users’ collaboration, so as to avoid many complicated issues such as users’ privacy and security. 
\vspace{-2pt}
\subsection{Action Space }

At the beginning of each time slot, the agent needs to make the following decisions: 1) Whether to access the spectrum resource, 2) How much energy to be used for data transmission, and 3) How to split time slots for data transmission and energy harvesting. Specifically, for the agent $n$, the variables $U_s^n$, $p_s^n$, and $\alpha_n$ are the corresponding access action, power control action and time allocation action, and their corresponding sub-action spaces are $\mathcal{A}_u$, $\mathcal{A}_p$, and $\mathcal{A}_T$ respectively. 

To reduce the complexity of the DDPG based  multidimensional joint resource scheduling algorithm, we combine the access action $U_s^n$ and the power control action $p_s^n$. That is,   $p_s^n(t)=0$  means that the SSU $n$ does not access the spectrum channel at time slot $t$, and thus $U_s^n(t)=0$. Similarly,  $p_s^n(t)>0$ means the $n$th SSU access the channel, i.e., $U_s^n(t)=1$. Then it is natural to set $\textbf{a}_n(t)=[p_s^n,\alpha_n]$ as the action vector. The  sub-action space of time allocation action $\alpha_n(t)$ satisfies $\mathcal{A}_T= [0,1]$. However, for the power control action $p_s^n$,  the maximum value of $p_s^n$ is fluctuating because the energy stored in its battery is dynamically changing, which can be expressed as 
\begin{equation}
\max\{p_s^{n}\}=\min\{\dfrac{E_n}{\alpha_nT}, \quad p_s^{max}\}.
\end{equation}
Thus the sub-action space satisfies $\mathcal{A}_p\in[0, \min(E_{max}/(\alpha_nT),p_s^{max})]$, where $\alpha_n\in[0,1]$. However, the fluctuating action space will lead to an unstable performance of the DRL. Therefore we normalize the power action as $\mathcal{A}_p=[0,1]$ and the action vector can be  expressed as 
\begin{equation}
\mathbf{a}_n(t)=[p_n,\quad \alpha_n],
\end{equation}
where power action $p_n\in\mathcal{A}_p$, and the corresponding transmit power is
\begin{equation}
p_s^n=p_n \text{min}\{p_{max},\dfrac{E_n}{\alpha_nT}\}.
\label{p_s}
\end{equation}
Obviously, such an action setting can guarantee the constraints on the transmit power in $(P_b)$ and on the time sharing factor in $(P_c)$  in the optimization problem.

\subsection{Reward Function Design}
In the DRL algorithm, the action of each agent is driven by rewards, which means that designing a reasonable reward function is crucial for the DRL algorithm. Considering the objective function  ($\mathbf{P1}$) and the constraint ($P_a$) - ($P_e$) in the optimization problem, we need to consider the following aspects when designing the sub-reward function.

\begin{itemize}
\item  As the objective function of the optimization problem  ($\mathbf{P1}$) is to minimize  packet losses, it is natural to set  $L_n(t)$ as the sub-reward for agent $n$, i.e.,
\begin{equation}
R_b=-L_n(t).
\end{equation}
Therefore, the more packets lost,  the greater the penalty will be obtained.

\item Considering the constraints of the SSU's QoS requirement in the optimization problem, the sub-reward function is designed as
\begin{equation}
R_r=
\begin{cases}
R_s^n(t),& R_s^n\geq R^0 \\
0,&\text{otherwise}
\end{cases}.
\label{r_f}
\end{equation}
Obviously the sub-reward function $R_r$ can motivate the agent to achieve a higher data rate, which also implies less loss of packets

\item In the EH-CR-NOMA IoT system,  the SSUs are not allowed to cause harmful interference to the PUs. Thus the interference to the PUs needs to be taken into account in the reward function.  Based on the constraints of PU's QoS requirement in the optimization problem, the sub-reward function can be designed as follows
 \begin{equation}
R_p=-\sum_{m=1}^{M}O_p^m(t),
\end{equation}
where we use $O_p^m(t)=1$ to indicate that the $m$-th PU is connected to the network and successfully communicated, and $O_p^m(t)=1$ otherwise.

\end{itemize}

Taking all the above factors into consideration, we designed the reward function of the system as
\begin{equation}
R_n(t)=w_1R_b+w_2R_r+w_3R_p,
\label{reward_all}
\end{equation}
where $w_1$, $w_2$ and $w_3$ are the nonnegative weighted factors.

\section{Aappdg Based Decentralized  Resource Management }
In this section, we begin with a brief introduction to the DRL technique. Then, the proposed AADDPG-based multidimensional resource management algorithm will be described in detail.
\vspace{-2pt}

\subsection{Basics for Deep Reinforcement Learning}

The DRL technique consists of five key components, including the agent and the system environment, in addition to the state space $\mathcal{S}$, action space $\mathcal{A}$, and reward functions $R$. The interrelationship between them can be described as a Markov Decision Process (MDP). During time step $t$ in each episode, each agent observes the system environment to obtain the current state $\mathbf{s}(t)$, and selects the appropriate action $\mathbf{a}(t)$ from the action space. After the action is executed,  the environment enters the next state  $\mathbf{s}(t+1)$, and the agent obtains the corresponding reward $R(t)$.  After designing these five elements rationally, the agent can learn the optimal strategy $\pi^*$ through continuous trial and error with the environment, i.e.,
$\pi^*:\mathcal{S}\longrightarrow\mathcal{A}$, which can maximize the accumulated reward, i.e., $ \sum_{t=1}^{\infty}\gamma^{t-1}R(t)$,
where $\gamma$ is reward discount parameter.

Based on the continuity of the action space $\mathcal{A}$ that can be handled, DRLs can be divided into value-based DRL, such as Deep Q network (DQN), Q learning, etc., and policy-based DRL, such as the policy gradient algorithm (PG) \cite{survey_drl}.  In the Q learning and DQN  \cite{dqn} algorithms, the action space is discrete, and the Q-value function 	$Q_{\pi}( \mathbf{s}, \mathbf{a};\bm{\theta})$ is introduced to evaluate each state-action pair under  policy $\pi$, where $\bm{\theta}$ is the parameters of the deep neural network (DNN). Driven by the reward function, after repeated exploration and training, the agent will learn the optimal policy $\pi^*$ with parameter $\bm{\theta^*}$. Based on this $\pi^*$, for any state, the action with the largest $Q_{\pi^*}(s,a;\bm{\theta^*})$ value will be selected and executed.  

For dynamic decision problems with continuous action space, policy-based DRL algorithms are good choices. Unlike the value-based DRL algorithm, a policy-based PG algorithm selects behavior based on the probability distribution. However, its network parameters are updated at the end of each episode, rather than at every step, which slows  convergence. 

The actor-critic (AC)  algorithm combines the advantages of Q-learning and PG algorithms \cite{DRLres2}, it can perform action selection in a continuous action space like PG, and  achieve a single-step update of network parameters like the Q-learning algorithm.   In each step $t$, the actor  selects the action based on the probability distribution $\pi_{\bm{\theta}}(\mathbf{a}\mid \mathbf{s})$, and the  critic evaluates the selected action based on the state-action value function $Q_{\pi_{\bm{\theta}}}(\mathbf{s},\mathbf{a})$. Then, the actor  modifies the probability of the action based on the critic's evaluation. The critic provides a temporal-difference (TD) error term in the
policy gradient to guide the actor’s learning as follows:

\begin{equation}
\triangledown_{\bm{\theta}}J(\bm{\theta})=\mathbb{E}_{\pi}[\triangledown_{\bm{\theta}}\log\pi_{\bm{\theta}}(\mathbf{s},\mathbf{a})\delta],
\end{equation}
where $\delta=R(t+1)+\gamma Q_\pi(\mathbf{s}(t+1),\mathbf{a}(t+1))-Q_\pi(\mathbf{s}(t),\mathbf{a}(t))$ is the TD error. The convergence of the AC algorithm  depends on the  Critic   network. However, since both actor and critic networks are updated in a continuous state, the data before and after the update has a large correlation, which causes the critic network  difficult to converge. Based on this, the Google DeepMind team has proposed  the DDPG, which has a structure like AC, but instead of outputting probabilities of behaviors, it outputs a specific action for continuous action prediction. It combines the DQN structure to improve the stability and convergence of AC.

\subsection{Multi-Agent DDPG with Action Adjuster }

\begin{figure*}[t]
	\centering
	\includegraphics[width=6in]{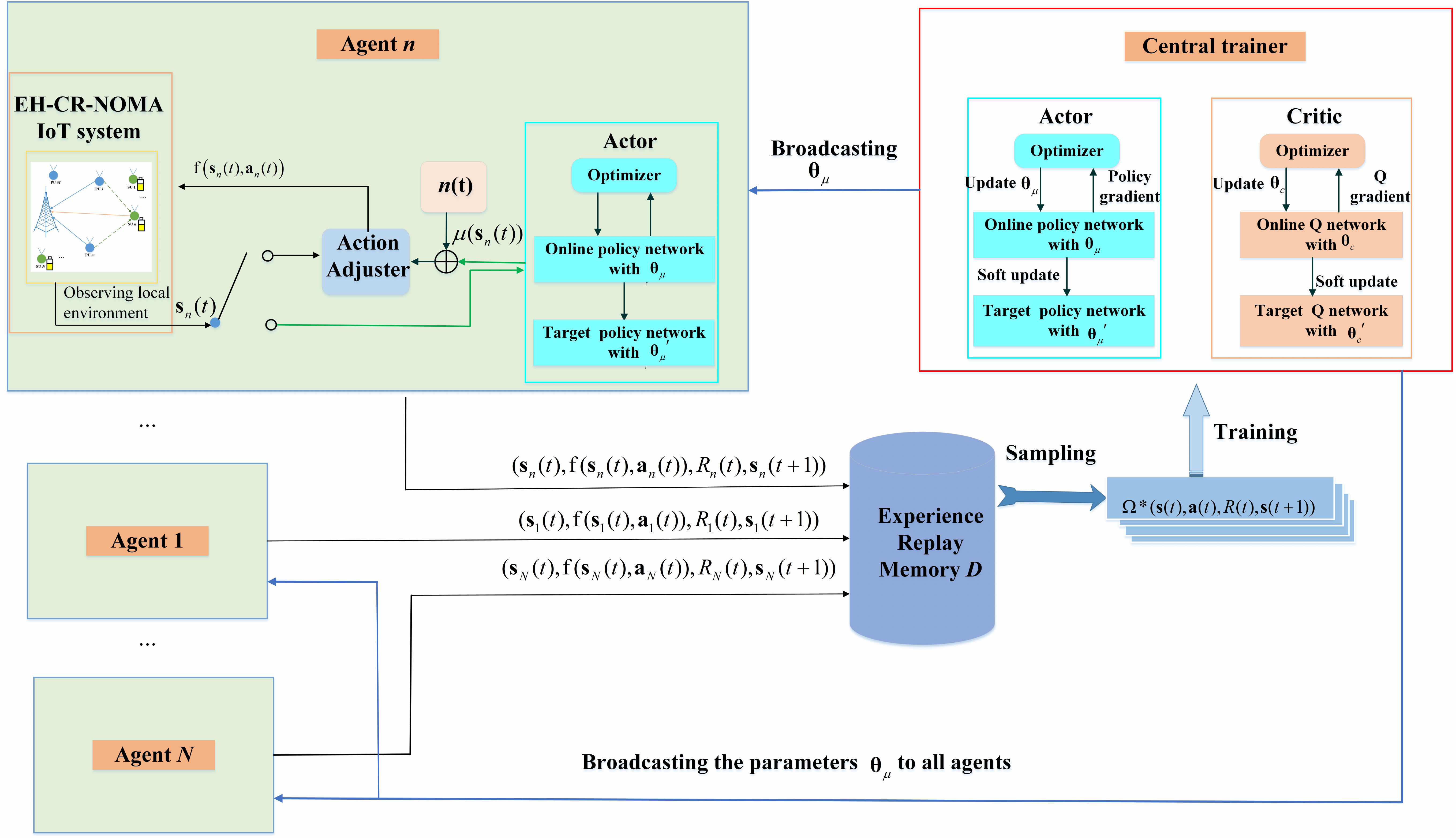}
	\caption{ The schematic framework of the proposed AADDPG algorithm.}
	\centering
	\label{ddpg}	
\end{figure*}
\vspace{-2pt}

In this paper, the action space in the problem of dynamic management of energy and time resources is  continuous and multidimensional, which cannot be solved by DQN algorithm. In addition, the convergence performance of PG and AC algorithm limits their applications in complex wireless communication systems \cite{pg}. DDPG is a model-free, off-policy actor-critical  algorithm combining the advantages of DQN and AC \cite{lianxu_drl}, which can learn the optimal deterministic strategy in the dynamic decision problem with multidimensional continuous action space. Therefore, DDPG algorithm is selected for the multidimensional continuous resource scheduling for the EH-CR-NOMA IoT system. 

In DDPG algorithm, four neural networks are used in each agent $n$ to learn the decision policy, which are  listed as follows:

$\bullet$ The actor network (also known as the policy network), with parameters $\bm{\theta}_\mu$, provides the desired solution for the optimization problem. The role of the network is to define parameterized policy $\mu$, which can generate corresponding action $\mu(\mathbf{s}_n|\bm{\theta}_\mu)$ for the input state $\mathbf{s}_n$.

$\bullet$ The target actor network, with parameters $\bm{\theta}_{\mu }^{\textquotesingle}$, has the same  structure as the actor network. It takes $\mathbf{s}_n$ as input and output  a target action  $\mu^{\textquotesingle}(\mathbf{s}_n|\bm{\theta}_{\mu}^{\textquotesingle})$.

$\bullet$ The critic network (also known as the Q network),  with parameter $\bm{\theta}_c$,  is primarily responsible for evaluating the policy $\mu$ generated by actor network. It takes the state and action as the network inputs and outputs the corresponding state-action value function $Q(\mathbf{s}_n,\mathbf{a}_n|\bm{\theta}_c)$.

$\bullet$ The target critic network, with parameter $\bm{\theta}_{c}^{\textquotesingle }$, has the same structure as the critic network. It is mainly in charge of calculating the target state-action value $Q^{\textquotesingle}(\mathbf{s}_n,\mathbf{a}_n|\bm{\theta}_{c}^{\textquotesingle })$.

The actor and critic target networks update their parameters through soft updating as follows
\begin{equation}
\begin{aligned}
\bm{\theta}_{\mu}^{\textquotesingle }(t)&=\tau \bm{\theta}_{\mu }(t)+(1-\tau)\bm{\theta}_{\mu}^{\textquotesingle}(t),\\
\bm{\theta}_{c}^{\textquotesingle }(t)&=\tau \bm{\theta}_{c }(t)+(1-\tau)\bm{\theta}_{c}^{\textquotesingle }(t),
\end{aligned}
\label{soft}
\end{equation}
where $0<\tau\ll1$ is the soft updating factor. Soft update removes the instability of the AC algorithm and speeds up its convergence.
These four networks can be divided into  two part: actor and critic. In actor part,   the  deterministic parameterized policy $\mu$ is the output, and the action can be expressed as 
\begin{equation}
\mathbf{a}(t)=\mu(\mathbf{s}_n(t)|\bm{\theta}_\mu(t))+n(t)
\label{act}
\end{equation} 
where $n(t)$ is the exploration noise, which is introduced to fully explore the action space \cite{duibi_ding}. It can be obtained by
\begin{equation}
n(t)=\begin{cases}
n^0(t)-t*\sigma ,& n(t)>n^{e} \\
n^{e},&\text{otherwise}
\end{cases}\label{nnn},
\end{equation}
where $n^0$, $n_e$, and $\sigma$ represent the maximum exploration noise, the minimum exploration noise, and the decreasing factor of the exploration noise, respectively, and these parameters can be adjusted according to the specific application system.

In addition, in order to prolong the battery life and improve the learning efficiency of the DRL-based algorithm, we design an action adjuster, the principle of which is shown in \textbf{Algorithm} \ref{aa}. At  time slot $t$, the agent $n$ observes the environment and obtains the state $\mathbf{s}_n(t)$. First of all, it needs to determine whether the remaining energy of the battery exceeds the  protection threshold $E^0$. If the electric quantity is too low, the actor and Critic networks would not be started, and the action with $p_s^n(t)= \alpha_s^n(t)=0$ would be output directly  to prevent the battery from over discharge. Besides,  to improve the learning efficiency and performance of the proposed algorithm, we adjusted the output action of actor network  with very low computational complexity as follows:
\begin{algorithm}[!t] 
	\caption{Principle of Action Adjuster} 
	\label{aa} 
	\begin{algorithmic}[1] 
		\REQUIRE  
		current environment state $\mathbf{s}_n(t)$
		\ENSURE 
		the real action vector $(p_s^n(t),\alpha_n(t))$
		\IF{$E_n(t)<E^0$ }
		\STATE {Input the state vector $\mathbf{s}_n(t)$ to the action adjuster, not to the Actor network};
		\STATE {Set the action vectors as $(p_s^n(t)=0,\alpha_n(t)=0)$ to prevent overdischarge of the battery}.
		\ELSE
		\STATE {Input the state vector $\mathbf{s}_n(t)$ to the Actor network and output the $\mu(\mathbf{s}_n(t);\bm{\theta}_\mu)$};
		\STATE {Input the action  $\mu(\mathbf{s}_n(t);\bm{\theta}_\mu)+n(t)$ to action adjuster};
		\STATE {Adjuster the action $p_{n}(t),\alpha_n(t)$  based on (\ref{ad_p})}.
		\ENDIF
		\STATE{Calculate the actual action by (\ref{action2}) and perform it to the EH-CR-NOMA IoT system}.		
	\end{algorithmic}
\end{algorithm}
\begin{equation}
\bar{\mathbf{a}}_n(t) =
\begin{cases}
(0,0),&  \text{if}\quad\log_2(1+\dfrac{p_s^n|h_s^n|^2}{\sigma^2})<R^0 \\
(p_n(t),\alpha_n(t)),&\text{otherwise}
\end{cases}.
\label{ad_p}
\end{equation}	
To reduce the computational complexity of action adjuster and to reduce the additional overhead, the interference from other users is not computed in Eq. (\ref{ad_p}). It implies that if the channel is used exclusively by SSU $n$, and it still fails to decode, then power percentage $p_n(t)$  and time sharing factor $\alpha_n(t)$ will be adjusted to 0. Otherwise, this communication will fail and the energy used for communication will be wasted. Hence, the  action adjuster can be defined as
\begin{equation}
\text{f}(\mathbf{s}_n(t),\mathbf{a}_n(t))=
\begin{cases}
	\bar{\mathbf{a}}_n(t), &\text{if}\quad E_n(t)\geq E^0 \\
(0,0 ),&\text{otherwise}
\end{cases},
\label{action1}
\end{equation}
where $\text{f}(\cdotp)$ is the mapping function of action adjuster. Since the transmit power in action $\text{f}(\mathbf{s}_n(t),\mathbf{a}_n(t))$ is normalized, if $ \text{f} (\mathbf{s}_n(t),\mathbf{a}_n(t))\neq(0,0)$  the following conversions are required before the action is executed

\begin{equation}
 \alpha_n=\mathcal{F}[2],\quad p_s^n(t)= \mathcal{F}[1]\text{min}\{p_{max},\dfrac{E_n(t)}{\alpha_n T}\},
 \label{action2} 
\end{equation}
where $\mathcal{F}[1]$ and $\mathcal{F}[2]$ respectively represent the normalized power and time-sharing factor output of the action adjuster.
We define the DDPG algorithm with the introduction of the action adjuster as the AADDPG algorithm.

The parameterized policy $\mu_{\bm{\theta}}(t)$ with action adjuster will be evaluated and criticized by the state-value function $Q(\mathbf{s}_n,\mathbf{a}_n)$ in the critic network. And the the actor network will be trained by maximizing the state-value function as 
\begin{equation}
	\mathbf{a}^*(\mathbf{s}_n)=\arg\max_\mathbf{a}Q(\mathbf{s}_n,\mathbf{a}|\bm{\theta}_c).
\label{a_star}
\end{equation}

\subsection{Centralized Training and Distributed Execution}
 Considering the limited computational power and energy resources of each SSU, together with overcoming the instability of the learning process in the multi-agent DRL algorithm,  we adopted the centralized training, distributed execution method as shown in Fig. \ref{ddpg}.  Specifically, at each time slot $t$,  as an independent agent, each SSU is configured with an Actor network with the same parameters $\bm{\theta}_\mu$ which is broadcasted  by the central trainer. Each SSU inputs its observed local state information $\mathbf{s_n}$ into the Actor network and outputs it as the corresponding action $\mathbf{a_n}$. The experience $(\mathbf{s_n}(t),\mathbf{a_n}(t+1),R_n(t),\mathbf{s_n}(t+1))$ obtained from this exploration will be stored centrally in the memory $\mathcal{D}$ for the centralized DDPG networks training.  To sum up, the centralized trainer trains the single DDPG network with the experiences gathered from all distributed agents. After each training, each agent replicates this DDPG network and independently executes its own resource scheduling decisions in a distributed manner. The approach can  effectively improves the learning efficiency of DRL and reduce the computational resources. It is also similar to the parameter sharing concept which enables the agent to draw the advantage from fact that each agent can learn experience from others for faster convergence \cite{DRL_power}. It is worth noting that  the same DDGP network in each agent still allows different action among agents, since they perform the same DDPG with different input local states. Based on the powerful learning capability of DRL and the adoption of such a centralized training and distributed execution strategy, each agent (SSU) can realize its own resource management without knowing the statistical information of the system (e.g., system state transfer probability, data arrival distribution function of SSUs, etc.) nor the information of other SSU users (e.g., channel, energy, transmit power, storage space, etc.).

Based on this centralized training model, in the rest of this section, we will omit the symbol $n$ that identifies the different agent and use $(\mathbf{s}(t),\mathbf{a}(t),R(t),\mathbf{s}(t+1))$ to characterize the information of the experience collected at the $t$-th time slot, which is stored in the experience memory $\mathcal{D}$ for networks training. Specifically, The critic network samples a $\Omega$-size mini-batch $ \{\mathbf{s}(i),\bar{\mathbf{a}}(i),R(i),\mathbf{s}(i+1)\}(i=1,\cdots,\Omega)$randomly from $\mathcal{D}$ to calculate the loss function of the critic network as
\begin{equation}
L(\bm{\theta}_c)=\dfrac{1}{\Omega}\sum_{i=1}^{\Omega}
[Q^{t}(i)-Q(\mathbf{s}(i),\text{f}(\mathbf{s}(i),\mu(\mathbf{s}(i)|\bm{\theta}_\mu))|\bm{\theta}_c(i))]^2,
\label{lossQ}
\end{equation}
where $Q^{t}$ is the target value of the state-value function, which can be calculated as
\begin{equation}
Q^{t}(i)=R(i)+\gamma Q^{\textquotesingle}\left(\mathbf{s}(i+1),\mu^{\textquotesingle}(\mathbf{s}(i+1)|\bm{\theta}_{\mu }^{\textquotesingle}(i))|\bm{\theta}_{c}^{\textquotesingle}(i)\right).
\end{equation}
Then the parameters of  the critic network can be updated by the gradient descent method as
\begin{equation}
\bm{\theta}_c\leftarrow \bm{\theta}_c-\beta_c\triangledown_{\bm{\theta}_c}L(\bm{\theta}_c),
\end{equation}
where $\beta_c$ denotes the learning rate of the critic network.
\begin{algorithm}[!t] 
	\caption{Training Process of AADDPG Based  Multidimensional Resource Management Algorithm} 
	\label{alg} 
	\begin{algorithmic}[1] 
		\REQUIRE  
		The learning rate $\beta_\mu$ and $\beta_c$, the soft update coefficient $\tau$, the discount parameter $\gamma$,   the mini-batch size $\Omega$. The channel information $h_s^n$, $\mathbf{G}_n$, and the communication status of all PUs $\bm{U}_p$.		
		\ENSURE 
		The optimal  policy  $\mu^*(s|\bm{\theta}_\mu^*)$.
		\STATE \textbf{Initialize :} Empty the relay memory $\mathcal{D}$.  Initialize the actor network $\mu(s|\bm{\theta}_\mu)$ and the critic network $Q(s,a|\bm{\theta}_c)$ of all agents with random parameters. Initialize the central trainer. Initialize serial number of transition $d = 0$.
		\FOR{each episode $k=1,\cdots,K^{all}$}
		\STATE{Reset the EH-CR-NOMA IoT   system;}
		\STATE{Randomly generate the battery information $E_n(0)$ and buffer information $B_n(0)$, and combine   $h_s^n(0)$, $\mathbf{U}_p(0)$ and $\mathbf{G}(0)$  to obtain the initial state $\mathbf{s}_n(0)$ for each agent;}
		\FOR {$t=1,\ldots,T^{all}$}
		\STATE{All agents copy the network  parameters $\bm{\theta}_\mu$ of the central trainer and updates their local networks.}
		\FOR{$n=1,\ldots,N$}
		\STATE{Observe and obtain the  state $\mathbf{s}_n(t)$ based on (\ref{state})};
		\IF{$t>1$}
		\STATE{Store the transition $\left({\mathbf{s}_n(t-1),\mathbf{a}_n(t-1),R_n(t-1),\mathbf{s}_n(t)}\right)$ into Memory $\mathcal{D}$};
		\STATE{$d=d+1$.}
		\ENDIF		
		\IF{$E_n(t)<E^0$}
		\STATE{Set $p_s^n(t)=\alpha_n(t)=0$};
		\ELSE
		\STATE {Take action $\mu(\mathbf{s}_n(t);\bm{\theta}_\mu)+n(t)$};
		\STATE{Transfer action to the action adjuster for action optimization, and adjust it to the actual action  according to (\ref{action1}) and (\ref{action2}).}
		\ENDIF
		\ENDFOR	
		\STATE{Perform the actual action in the EH-CR-NOMA system};
		\STATE{Calculate the reward $R_n(t)$ according to (\ref{reward_all}) for all agents};
		\STATE {Soft update the target networks of central trainer according to (\ref{soft})};
		\IF{$d\geq\dfrac{1}{3}\mid\mathcal{D}\mid$}
		\STATE{Sample a minibatch with $\Omega$ transitions from $\mathcal{D}$};
		\STATE{Update $Q(s, a; \bm{\theta}_c)$ by minimizing the loss in (\ref{lossQ})};
		\STATE{Update the policy $\mu(s; \bm{\theta}_\mu)$ by maximizing the policy 	gradient in (\ref{gra})};
		\STATE{Update the exploration noise  $n(t)$ base (\ref{nnn})}.		
		\ENDIF
		\ENDFOR	
		\ENDFOR	
	\end{algorithmic}
\end{algorithm}
By using the parameters of the actor and  critic networks, the objective function of the maximization problem in (\ref{a_star}) can be rewritten as 
\begin{equation}
J(\bm{\theta}_\mu)=Q\left(\mathbf{s},\text{f}(\mathbf{s} ,\mu(\mathbf{s}|\bm{\theta}_\mu))|\bm{\theta}_c\right)\label{bbb}.
\end{equation}
The parameter $\bm{\theta}_\mu$ can be updated by maximizing the output of the critic network as in (\ref{bbb}). Hence, considering the fact that the state-action function is differentiable and the action space is continuous,  the actor network can be updated by the policy gradient with the ascent factor as 
\begin{equation}\
\begin{aligned}
\triangledown_{\bm{\theta}_\mu}J(\mu(\bm{\theta}_\mu))  =\dfrac{1}{\Omega}\sum_{i=1}^{\Omega}\triangledown_\mathbf{a}Q(\mathbf{s}(i),\mathbf{a})|\bm{\theta}_c)\triangledown_{\bm{\theta}_\mu}\text{f}(\mathbf{s},\mu(\mathbf{s}(i))|\bm{\theta}_\mu).
\end{aligned}
\label{gra}
\end{equation}
The specific training process is concluded in \textbf{Algorithm} \ref{alg}. Specifically, at each time slot $t$, agent $n$ observes local environment information to obtain state $\mathbf{s}_n(t)$. Based on this state  $\mathbf{s}_n(t)$ and the output of the local actor network, the action adjuster obtains action $\text{f}(\mathbf{s}_n(t),\mu(\mathbf{s}_n(t)|\bm{\theta}_\mu))$. After transforming it into $p_s^n(t)$ and $\alpha_n(t)$, agent $n$ will conduct data communication with power $p_s^n(t)$ for a time duration of  $\alpha_n(t)T$. After all agents have performed their actions to the EH-CR-NOMA IoT system, each agent will receive its own reward $R_n(t)$ and move to the next state $\mathbf{s}_n(t+1)$. The experience tuple $(\mathbf{s}_n(t), \text{f}(\mathbf{s}_n(t),\mu(\mathbf{s}_n(t)|\bm{\theta}_\mu)), R_n(t), \mathbf{s}_n(t+1))$ of all agent will be stored  together in memory $\mathcal{D}$. The central actor and critic network will be updated based on a mini-batch experience tuple randomly sampled from the $\mathcal{D}$. And the updated parameters will be  broadcast  to each agent by the central trainer at the begin of each time  slot. 
After training, the optimal multidimensional joint resource scheduling policy $\mu^*(\mathbf{s}|\bm{\theta}_\mu^*)$ for the EH-CR-NOMA IoT system will be learned, and the optimal network parameters $\bm{\theta}_\mu^*$  will be broadcasted to all agents for online application.
\section{Simulation Results and Discussions }

In this section, the performance of the proposed  AADDPG algorithm is evaluated by using computer simulation.
In addition, we compare the proposed AADDPG-based multidimensional resource  management algorithm with the following schemes.
\begin{itemize}
	\item Greedy algorithm: Each SSUs transmit data at the maximum power allowed, i.e., $p_s^n(t)=\text{min}\{p_{max},\dfrac{E_n}{\alpha_nT}\}$, where  $\alpha_n$ is calculated by AADDPG.
	\item  Random algorithm: Both the time sharing factor $\alpha_n$ and the transmit power percentage $p_n(t)$ are set to a random value between 0 and 1, respectively.	
	\item Constant $T$ + AADDPG $P$: The time sharing factor is set as $\alpha_n=0.5$ for all accessed SSUs, and $p_n(t)$ is obtained by the AADDPG;
	\item  Constant $T$+ Random $P$: Time sharing factor is set as $\alpha_n=0.5$ for all agents, and $p_n(t)$  is set to a random value between 0 and 1.
\end{itemize}

Unless stated explicitly, the simulation parameters are set as in Table \ref{Pa}. All results are obtained based on the deep learning framework in TensorFlow 1.14.0. 
\begin{table*}[!t]
	\small
	\caption{SIMULATION PARAMETER SETTING}
	\label{Pa}
	\begin{center}	
		\renewcommand\arraystretch{2} 
		\begin{tabular}{|l| l|l|l|}
			\hline
			\textbf{Parameters} &\textbf{Values}&\textbf{Parameters} &\textbf{Values}  \\	
			\hline	 
			Packet size $C$ (bits)& $2500$&Total number of episode $K^{all}$ &200\\
			soft Updating factor $\tau$&$0.01$&Total step in each episode $T^{all}$& 80\\
			Size of  mini-batch $\Omega$& 40 &Learning rate of the actor and critic  & $0.001,0.002$\\
			Transfer probabilities $P_1$ and $P_2$ &0.3, 0.8&SSUs distribution range  (m)&$15-60$\\
			Thresholds $R^0$  and $R^1$ $(bit/s/Hz)$  & 0.3, 0.3&	PUs distribution range  (m)&$5-20$\\
			Discount parameter $ \gamma$& 0.9&Reward weight $w_1$, $w_2$,  $w_3$&  $8/C$, $10$,  $4$ \\
			Noise power $\delta^2$ (dBmW)&-150&	Number of  neurons in hidden layers & 256, 256, 128\quad\\
		    Capacity of experience replay memory $\mid \mathcal{D}\mid$ &10000&Range of Bandwidth  ($K$ Hz) &$\{5-15\}$\\
			Energy harvesting coefficient $\eta$&0.9&	Range of SSUs numbers  &$\{1-50\}$\\	
			Transmit power of PUs $p_p$ (W)&1&Range of PUs numbers  &$\{1-10\}$\\
			Maximum transmitting power $p_{max}$(W)&0.5&The range of data arrival rate $\lambda$  &$\{0.5-2.5\}$\\
			Capacity of the battery $E_{max}(J)$ &2&Range of Buffer capacity $B_{max}$&$\{3-17\}$\\ 	
			\hline
		\end{tabular}
	\end{center}
\end{table*}
\subsection{Performance Verification of Action Adjusters and Access Mode }
\begin{figure}[t]
	\centering
	\subfigure[Experiment \uppercase\expandafter{\romannumeral1}  ]{\includegraphics[width=0.4\textwidth]{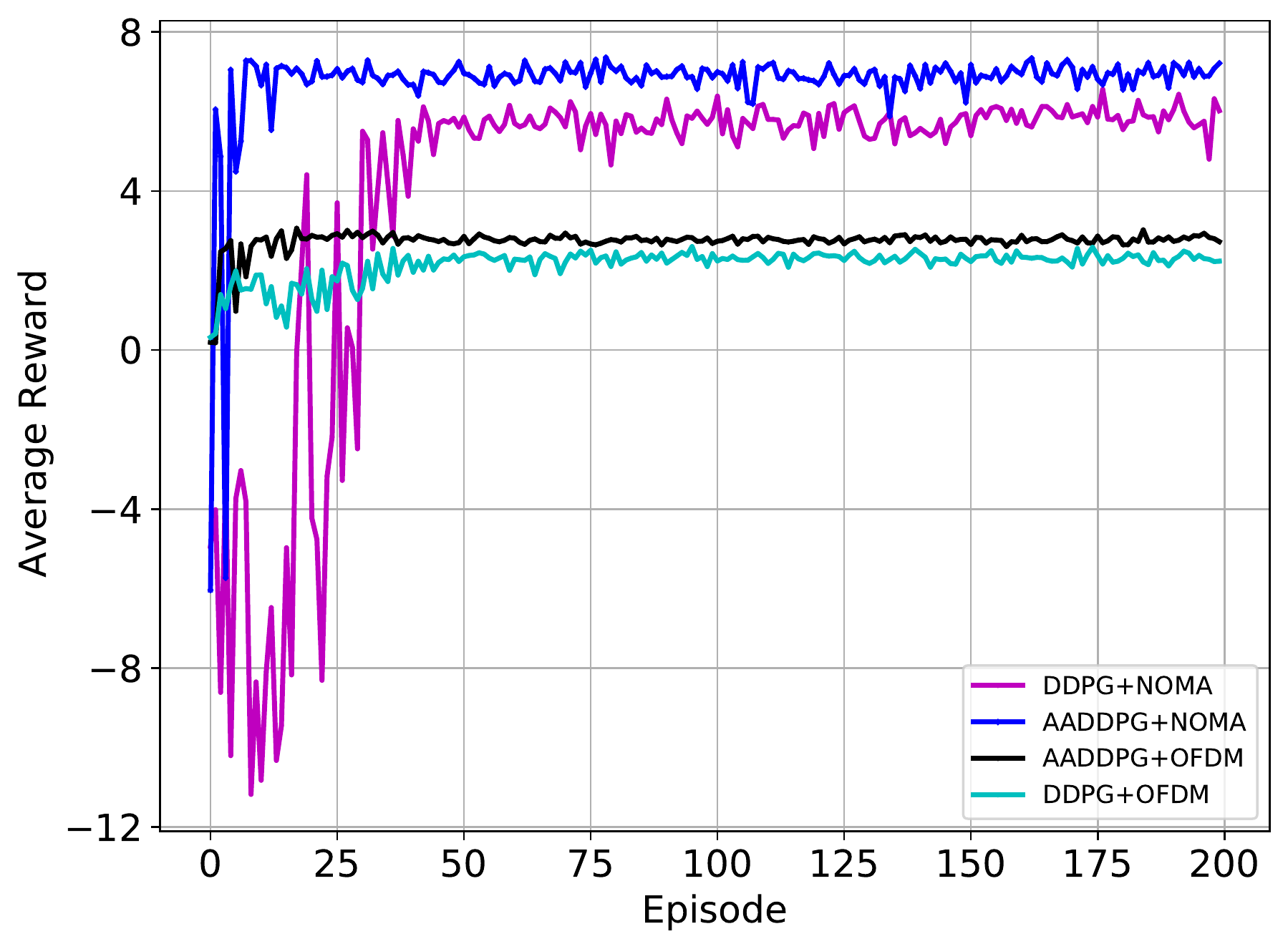}}\quad
	\subfigure[Experiment \uppercase\expandafter{\romannumeral2} ]{\includegraphics[width=0.4\textwidth]{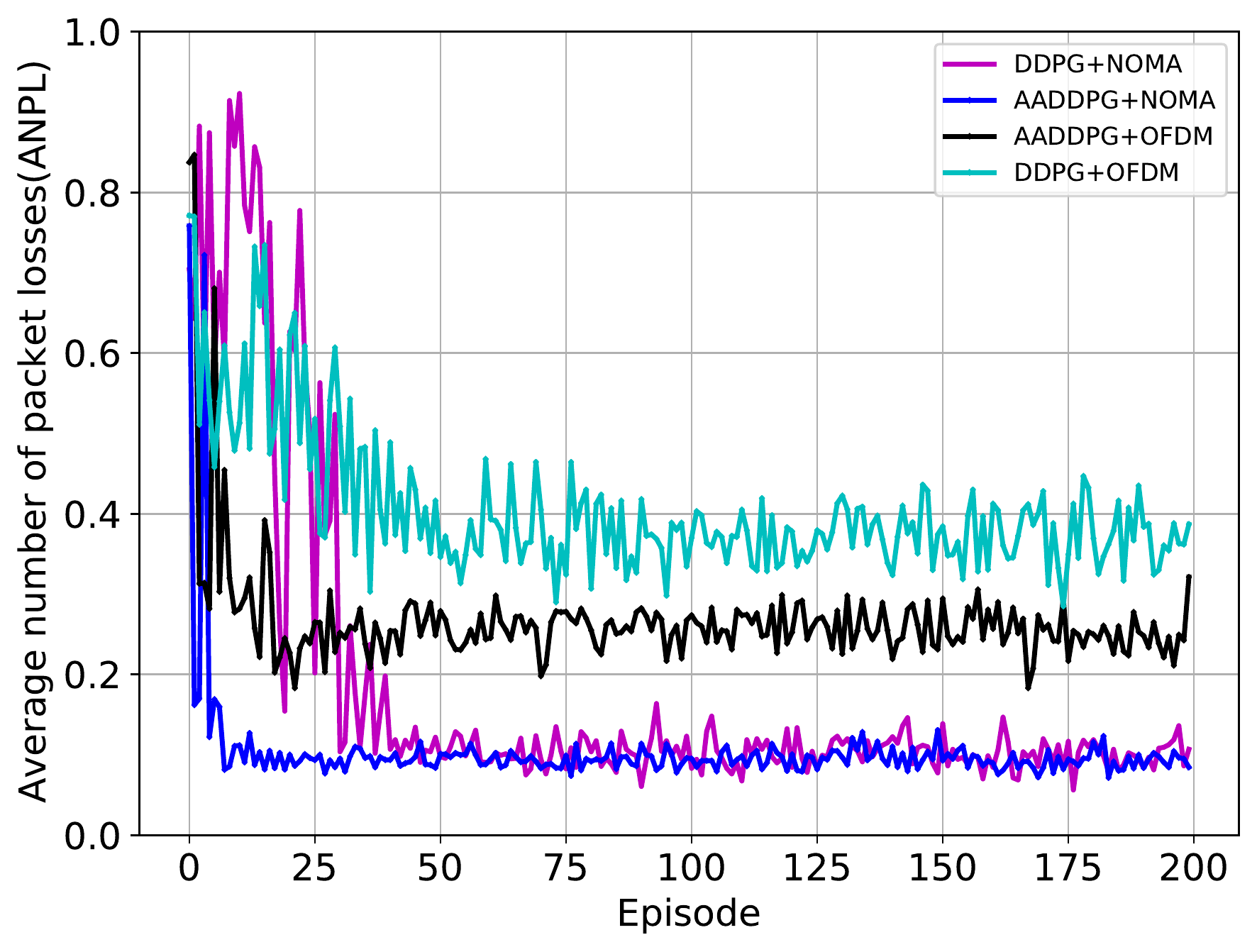}}\quad
	\subfigure[Experiment \uppercase\expandafter{\romannumeral3} ]{\includegraphics[width=0.4\textwidth]{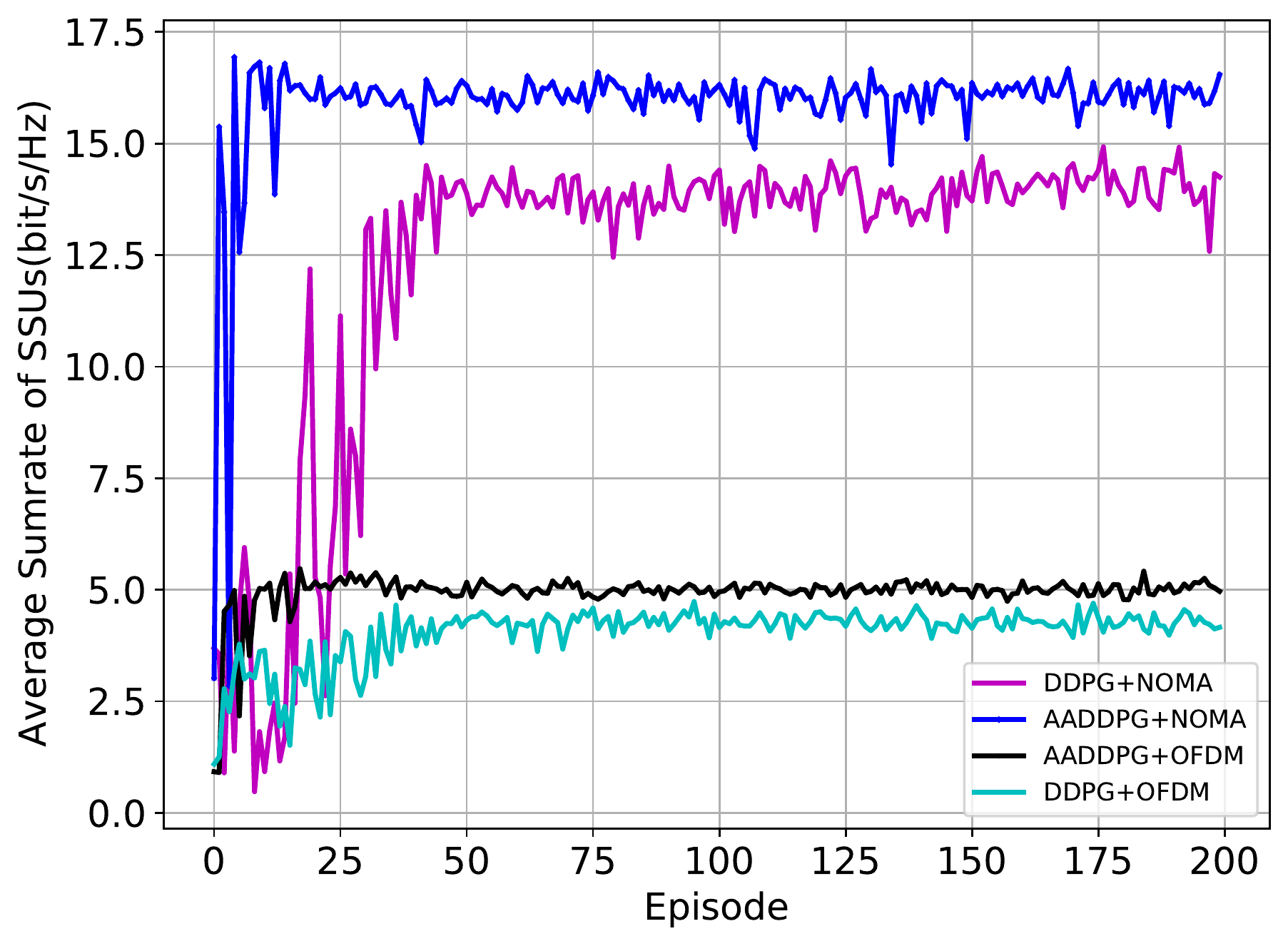}}
	\caption{Performance verification of action adjuster and access mode for EH-CR system, where $\lambda=1$, $B_{max}=6$, $M=3$, and $N=20$.}
	\label{aa_noma}
\end{figure}

First, we verify the effectiveness of the proposed action adjuster  and compare the  NOMA and OFDM access modes on the system performance.  Fig. \ref{aa_noma} shows the training effect against the training episodes in terms of the average reward per time slot per agent, the average number of loss packet  per time slot per agent, and the average sumrate of SSUs per time slot.  It should be noted that all algorithms in the figure are based on the DDPG framework designed in this paper, and the differences are mainly in the user's multiple access method and whether the action adjuster is used.

First, it can be seen that the performance of all algorithms improves  with the increase of the number of training episodes and eventually tends to be stable, which proves the convergence of the proposed algorithm.
Besides, the figure shows that the convergence speed of the algorithm is significantly accelerated after the introduction of the action adjuster in Algorithm \ref{aa}. AADDPG can achieve convergence in 5 episodes in both NOMA and OFDM access modes,  while DDPG takes about 37 episodes. In addition, it can be found that the performance is also improved after adding the action adjuster. In the  NOMA access mode, the average reward obtained by the AADDPG algorithm is 30\% higher than that of DDPG, the average sumrate of SSUs of AADDPG algorithm is about 15\% better than the DDPG. The  ANPL  performance of the DDPG algorithm is about 80\% worse than that of AADDPG in the OFDM access mode. These  results lead to the conclusion that the proposed action adjuster is very effective.

Fig. \ref{aa_noma} also verifies that the use of the NOMA mode in the constructed EH-CR  IoT system leads to better  performance than OFDM. Specifically, the reward value of AADDPG  in the NOMA  mode is about 2.5 times that in the OFDM mode,    ANPL  performance in the NOMA mode is about 3 times better than that in the OFDM mode, and the average sumrate of SSUs with NOMA mode in AADDPG is about 2.2 times higher that in OFDM.

\subsection{Comparison of Training Effects}
\begin{figure*}[!t]
	\setlength{\belowcaptionskip}{-2pt} 
	\centering
	\subfigure[Experiment \uppercase\expandafter{\romannumeral1} ]{\includegraphics[width=0.45\textwidth]{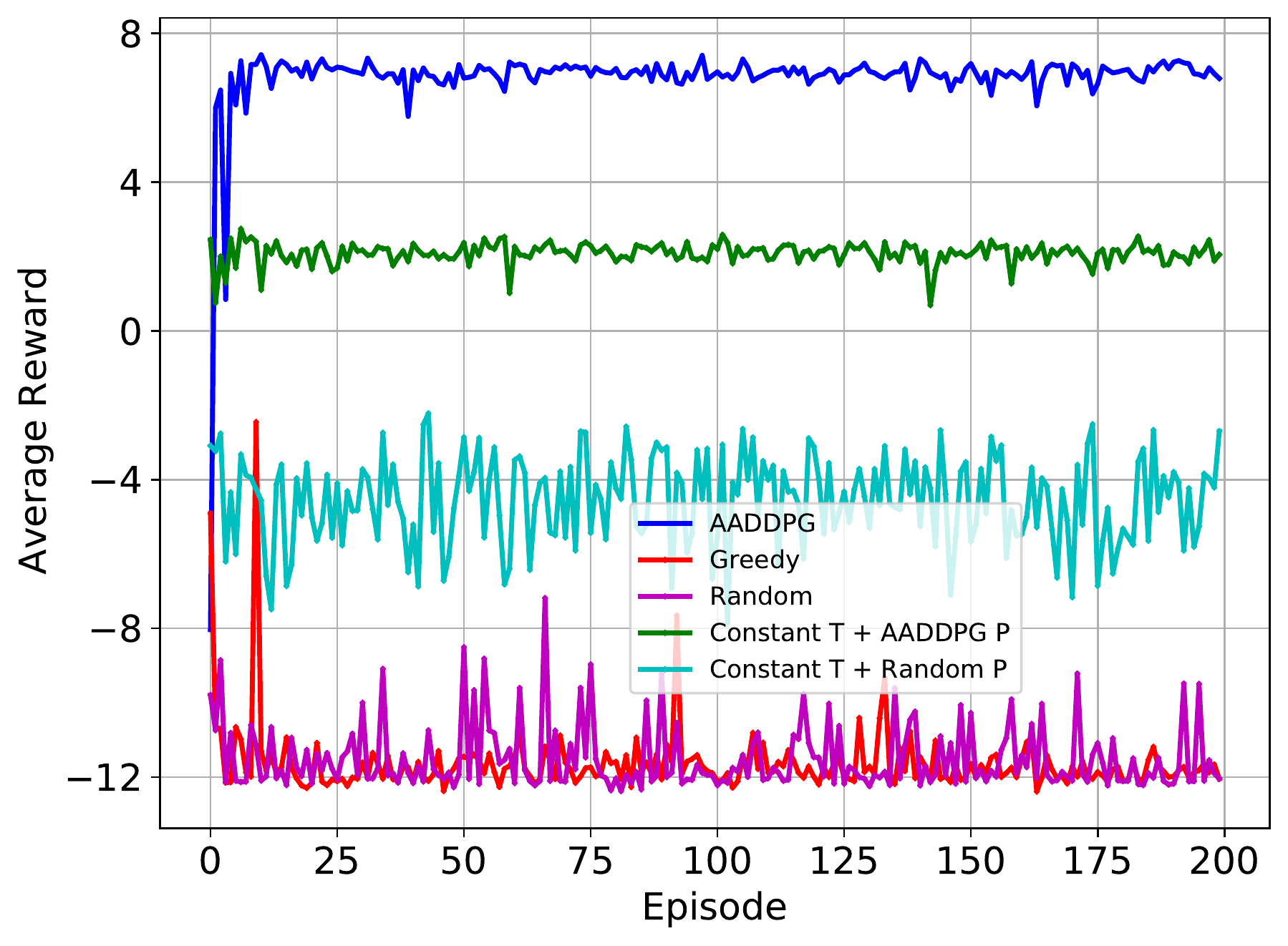}}\quad \quad
	\subfigure[Experiment \uppercase\expandafter{\romannumeral2} ]{\includegraphics[width=0.45\textwidth]{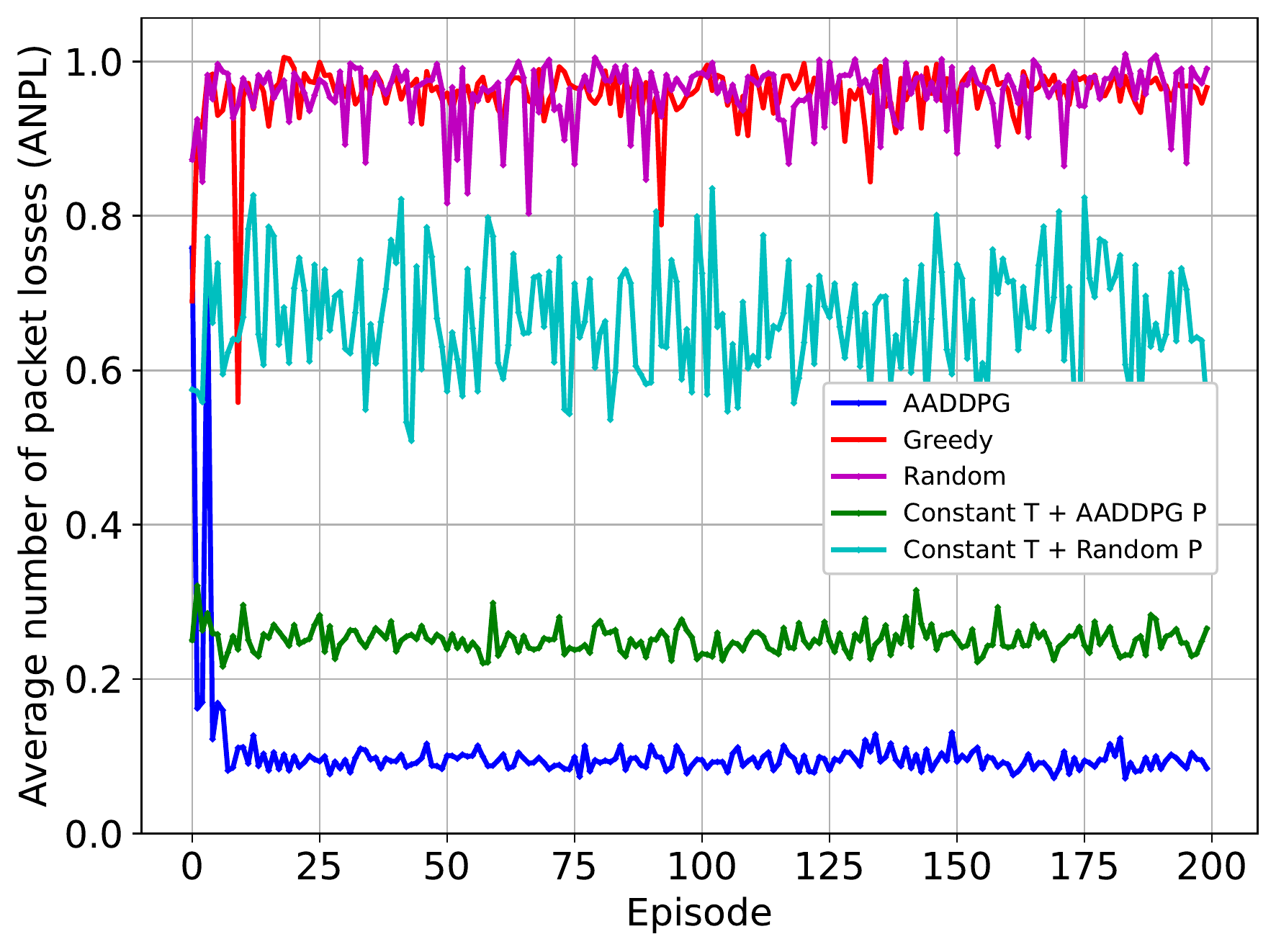}}\quad\quad
	\subfigure[Experiment \uppercase\expandafter{\romannumeral3} ]{\includegraphics[width=0.45\textwidth]{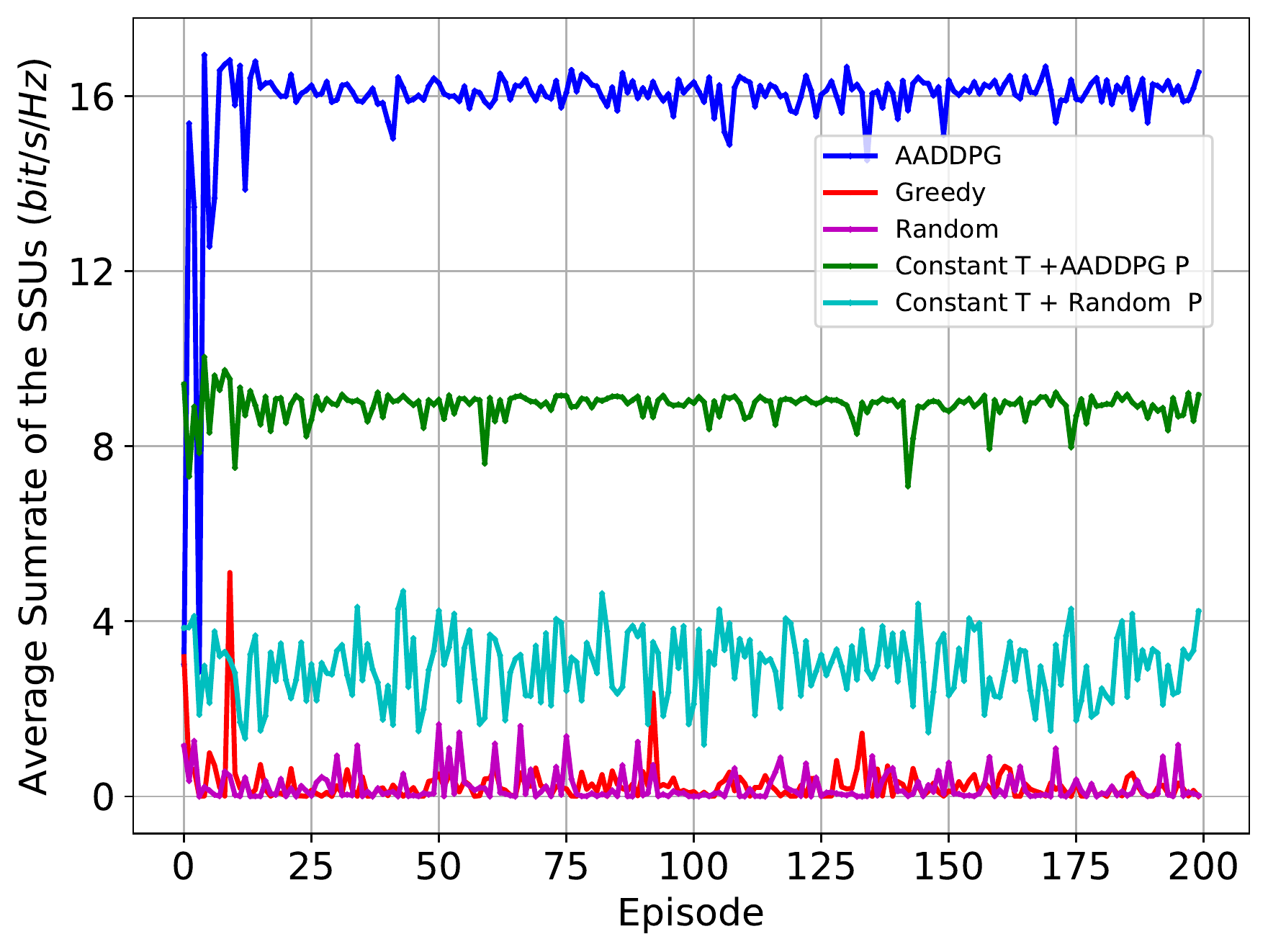}}\quad\quad
	\subfigure[Experiment \uppercase\expandafter{\romannumeral4} ]{\includegraphics[width=0.45\textwidth]{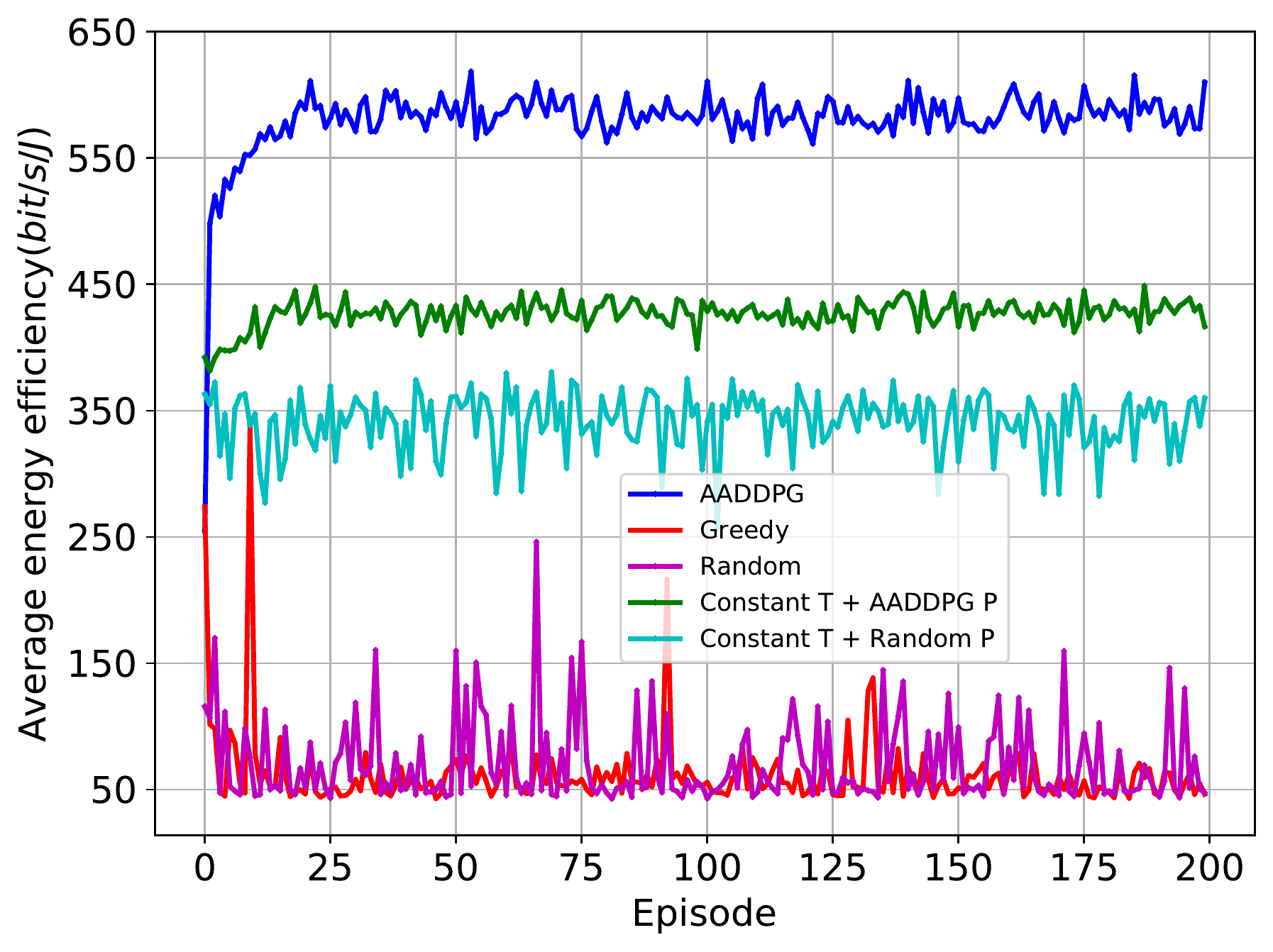}}
	\caption{Training effects comparison of different algorithms, where $\lambda=1$, $B_{max}=6$, $M=3$, and $N=20$.}
	\label{training_com}
\end{figure*}

To demonstrate the effectiveness of AADDPG, we compared its training effect with the other four benchmark algorithms in Fig. \ref{training_com}. It shows the training effect versus episode for different schemes, where the average reward and ANPL  are shown in Fig. \ref{training_com} (a) and (b), respectively. The average sumrate of all SSUs and the average energy efficiency, defined as $\left(\sum_{n=1}^{N}R_p^n(t)+\sum_{m=1}^{M}R_s^m(t)\right)/\sum_{m=1}^{M}U_p^m(t)p_p$ are depicted in Fig. \ref{training_com} (c) and  (d) respectively. 

As can be shown from Fig. \ref{training_com}, even for a single episode, the proposed AADDPG algorithm can already achieve better performance than the random and greedy algorithm. The performance gain of  AADDPG algorithm over the benchmark algorithms can be further improved by increasing the number of episodes from 1 to 5. The random and greedy schemes result in such poor performance because the decisions they are not based on a long-term goal. All SSUs in greedy algorithm transmit data at the current maximum battery power, which  leads to severe interference between users, further leading to data communication failure and an inefficient use of energy. In addition, it can also be found  that  except the AADDPG, the constant $T$ + AADDPG $P$ algorithm is able to obtain the best performance among the benchmark algorithms. This is due to the fact that the time sharing factor is set to a constant value of 0.5 in the constant $T$ + AADDPG $P$ algorithm, i.e., half of each time slot is used for data communication and the remaining half for energy harvesting, which is obviously a reasonable compromise, and the data transmit power in the constant $T$ + AADDPG $P$ algorithm is calculated by the AADDPG algorithm, which can further improve its performance, as evidenced by the performance comparison with the constant $T$ + Random $P$ algorithm. However,  as evidenced by these subfigures,  we can see that the proposed AADDPG scheme outperforms the constant $T$ + random $P$ and the other benchmark schemes. The AADDPG can always achieve a considerably better reward,  ANPL, sumrate, and  energy efficiency. Specifically, the average reward,  ANPL, sumrate of SSUs, and energy efficiency performance of the AADDPG algorithm are about 260\%, 160\%, and 200\%, and 30\% better than the constant $T$ + AADDPG $P$  algorithm, respectively.
 From the training results in Fig. \ref{aa_noma} and Fig. \ref{training_com}, it can be seen that the proposed AADDPG algorithm  can converge quickly and realize better system performance. 
 
 \subsection{The Relationship Between Average Delay and $B_{max}$}
 
In this part, we investigate the relationship between the $B_{max}$ and the average  access delay of all successful transmitted packets.
 
 \begin{figure}[!t]
 	\centering
 	\includegraphics[width=3.8in]{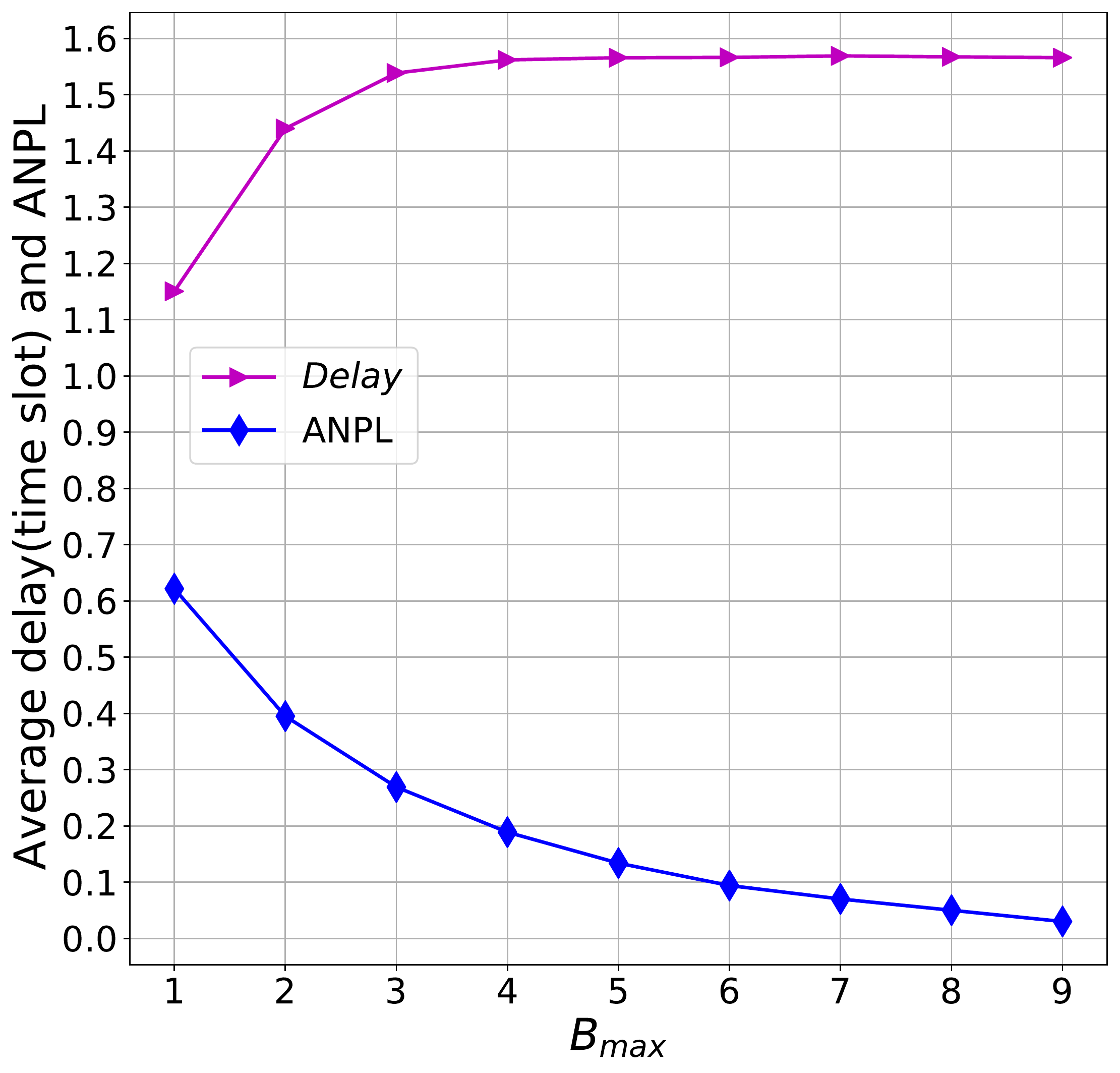}
 	\caption{Effect of $B_{max}$ on delay and ANPL with $\lambda=1$, $M=3$ and $N=20$.}
 	\centering
 	\label{delay}	
 \end{figure}

 Fig. \ref{delay} shows the average delay of successfully transmitted packets and the average number of lost packets in the proposed algorithm as $B_{max}$ increases. It can be found that the average number of packet loss is decreasing as $B_{max}$ increases. However, we notice that larger $B_{max}$ brings a relatively higher average delay. Specifically, when $B_{max}=1$, the system can obtain the lowest average delay of about $1.15$, and the average delay keeps increasing as $B_{max}$ increases. And when $B_{max}\geq 5$, the average delay  leveled off.
 
 The reason behind this interesting phenomenon is that when $B_{max}$  is very small, it means that less data is waiting to be transferred and there is no need to wait a long time in the queue, but at the cost of losing many packets because the buffer capacity is too small. The low latency in this case does not make any sense. As $B_{max}$ increases, the number of packets that can be stored in the buffer increases, i.e., the amount of time to wait before transmission also increases. However, as $B_{max}$ increases further and with the execution of the AADDPG algorithm, the ANPL drops to smaller values (between 0.04 and 0.13) and the average delay plateaus.

\subsection{Algorithm Performance Comparison}
 After training, the neural network in the central trainer will converge on the parameter $\bm{\theta}_\mu^*$, which will be broadcast to all the agents for real-time resource management. In this section, we will perform a performance test on the trained algorithm, where the data for the training and test sets are generated simultaneously and distributed in a $7:3$ ratio. We compare the performance of the test results with other benchmark algorithms. To better demonstrate the performance advantages of the proposed algorithm, the following two benchmark algorithms are introduced in the performance test comparison: namely DQN algorithm and the allocation algorithm based on request (AoR). In the DQN algorithm, both time and power resources are equally divided into 20 levels, i.e., 400 output dimensions for each agent. In the AOR, we adopt the resource allocation scheme in \cite{power_control}, which is implemented on the premise that all SSUs are informed in advance of the global information of the system, including all CSI, probabilistic statistics, arrival data information, buffer length, and collected energy, etc. 
 In the AoR, the $\alpha_n(t)=\dfrac{C*\lambda}{R^0*BW}$, where $C$ is the packet size, $\lambda$ is the  packets arrival rate, $R^0$ denotes the rate threshold of each SSU, and $BW$ is the bandwidth. That is, we set  $\alpha_n(t)$ to be a constant, which is the average time it takes to transmit the packets arriving in each time slot. The action $U_s^n$ and  $p_s^n$ will be calculated according to \textbf{Algorithm} \ref{aaa} as follows.	
 \begin{algorithm}[!t] 
 	\caption{Principle of AoR} 
 	\label{aaa} 
 	\begin{algorithmic}[1] 
 		\REQUIRE  
 		$M$, $N$, $\rho$, $R^0$, $h_p^m$, ($m\in\mathcal{M}$), $h_s^n$, $E_n(t)$,  ($n\in \mathcal{N}$)
 		\ENSURE 
 		the  action  $U_s^n$ and  $p_s^n$ for all SSUs.
 		\STATE \textbf{Initialize :} Initialize the access state of all SSUs as $U_s^n=0, n\in\mathcal{N}$; Initialize the transmit power as $p_s^n=0, n\in\mathcal{N}$
 		\STATE {Sort the SSUs with the increasing order of $|h_s^n|^2*E_n$, i.e., $\textbf{Index}=\text{argsort}\{|h_s^1|^2*E_1, |h_s^2|^2*E_2,\cdots,|h_s^n|^2*E_n\}$};
 		\FOR{$i$ in $\textbf{Index}$}
 		\STATE {Calculate the minimum transmit power required for the $i$th SSU $\textbf{Index}[i]$ in order to meet the decoding requirements, i.e.,  $p_{s,min}^{\textbf{Index}(i)}=\dfrac{2^{(R^0-1)\left(\sigma^2+\sum_{j=1}^{i-1}(U_s^{\textbf{Index}[j]}*p_s^{\textbf{Index}[j]})\right)}}{h_s^{\textbf{Index}[i]}}$}.
 		\IF{$\dfrac{E_n}{\alpha_n}\geq p_{s,min}$ }
 		\STATE {Set $p_s^{\textbf{Index}[i]}=p_{s,min}^{\textbf{Index}(i)}$};
 		\STATE {Set $U_s^{\textbf{Index}[i]}=1$};
 		\ENDIF
 		\ENDFOR	
 	\end{algorithmic}
 \end{algorithm}

\begin{figure}[t]
	\centering
	\includegraphics[width=4.5in]{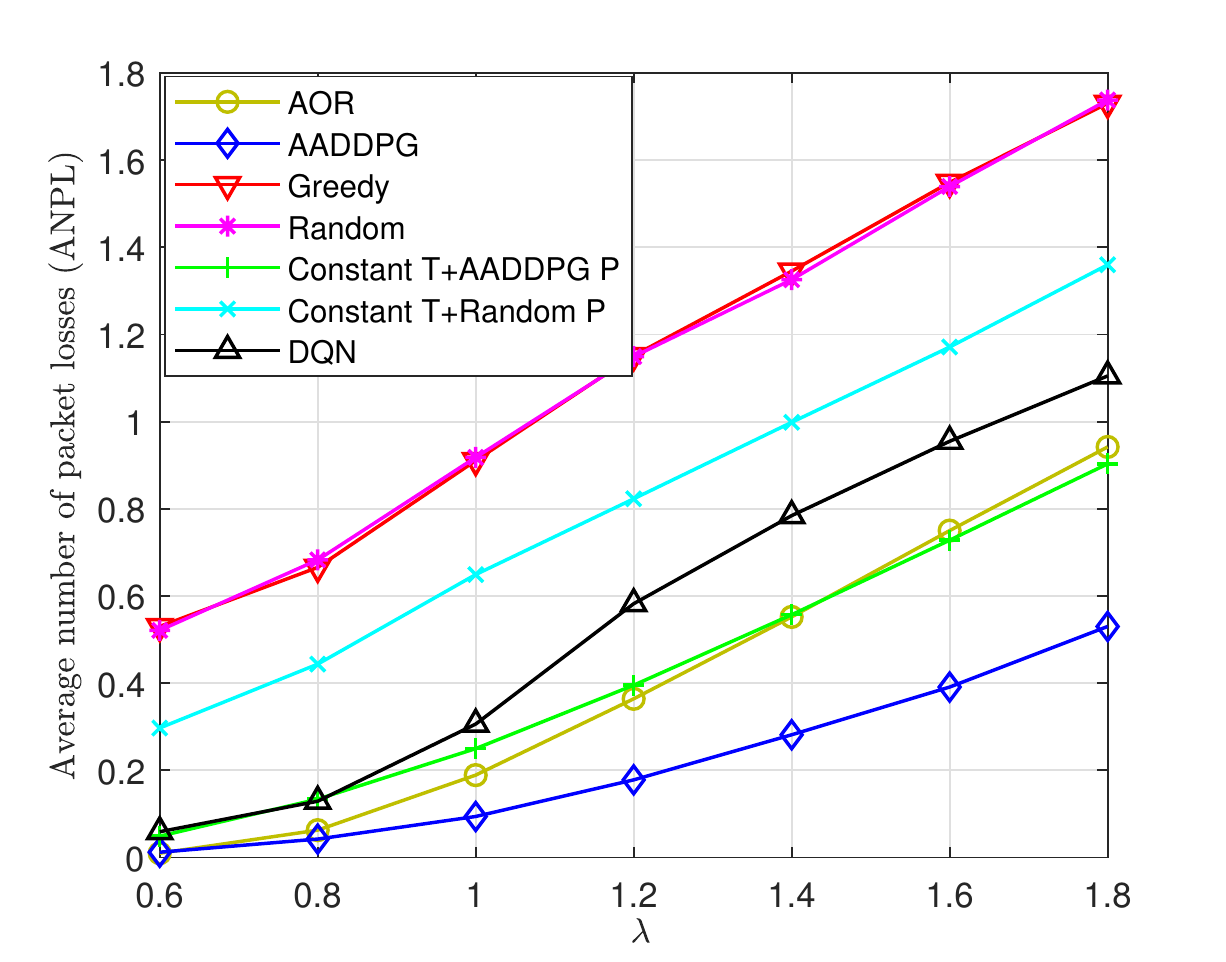}
	\caption{ Impact of $\lambda$ on the performance of the proposed AADDPG algorithm.}
	\centering
	\label{lam}	
\end{figure}

\begin{figure}[t]
	\centering
	\includegraphics[width=4.5in]{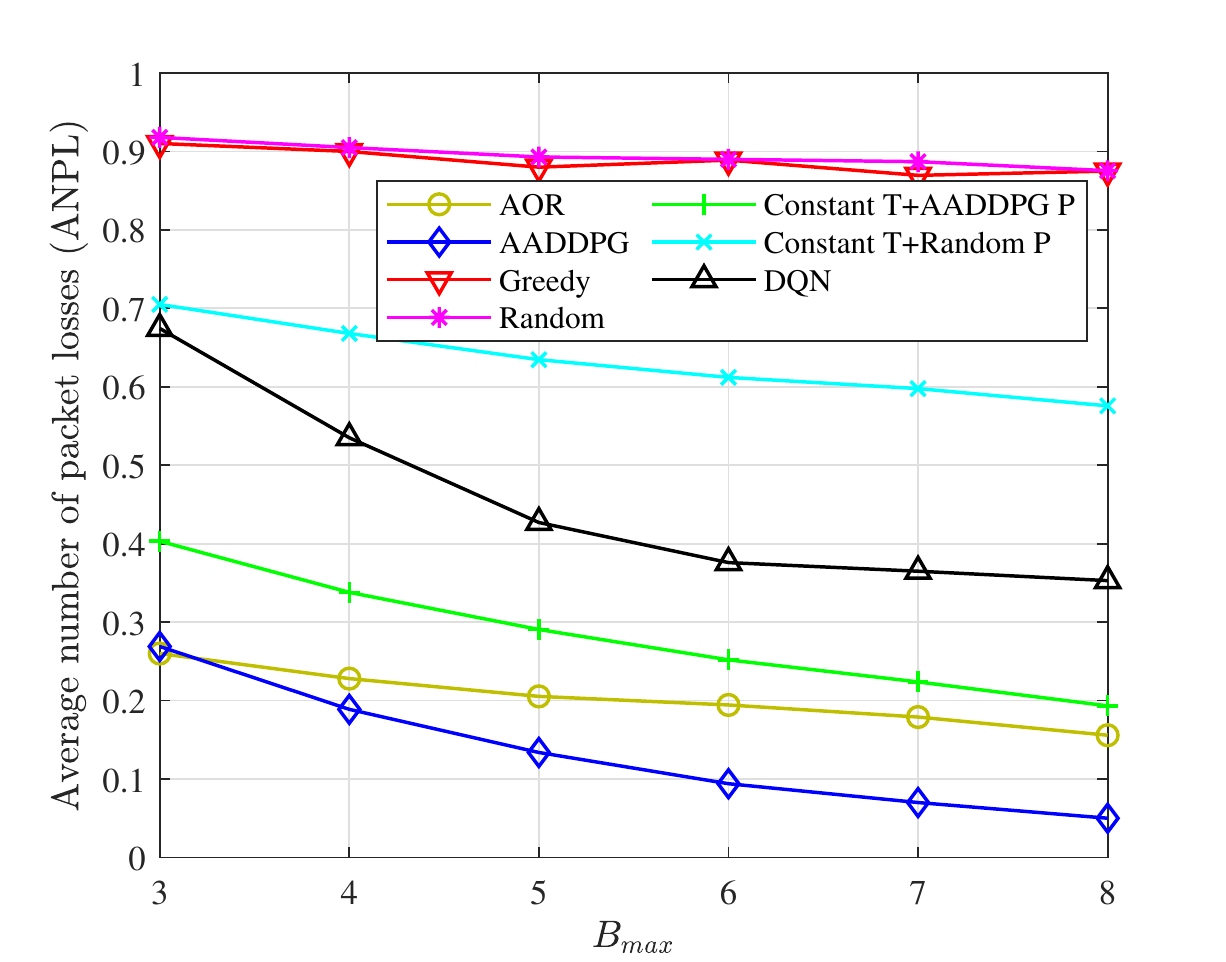}
	\caption{ Impact of $B_{max}$ on the performance of the proposed AADDPG algorithm.}
	\centering
	\label{bmax}	
\end{figure}

\begin{figure}[t]
	\centering
	\includegraphics[width=4.5in]{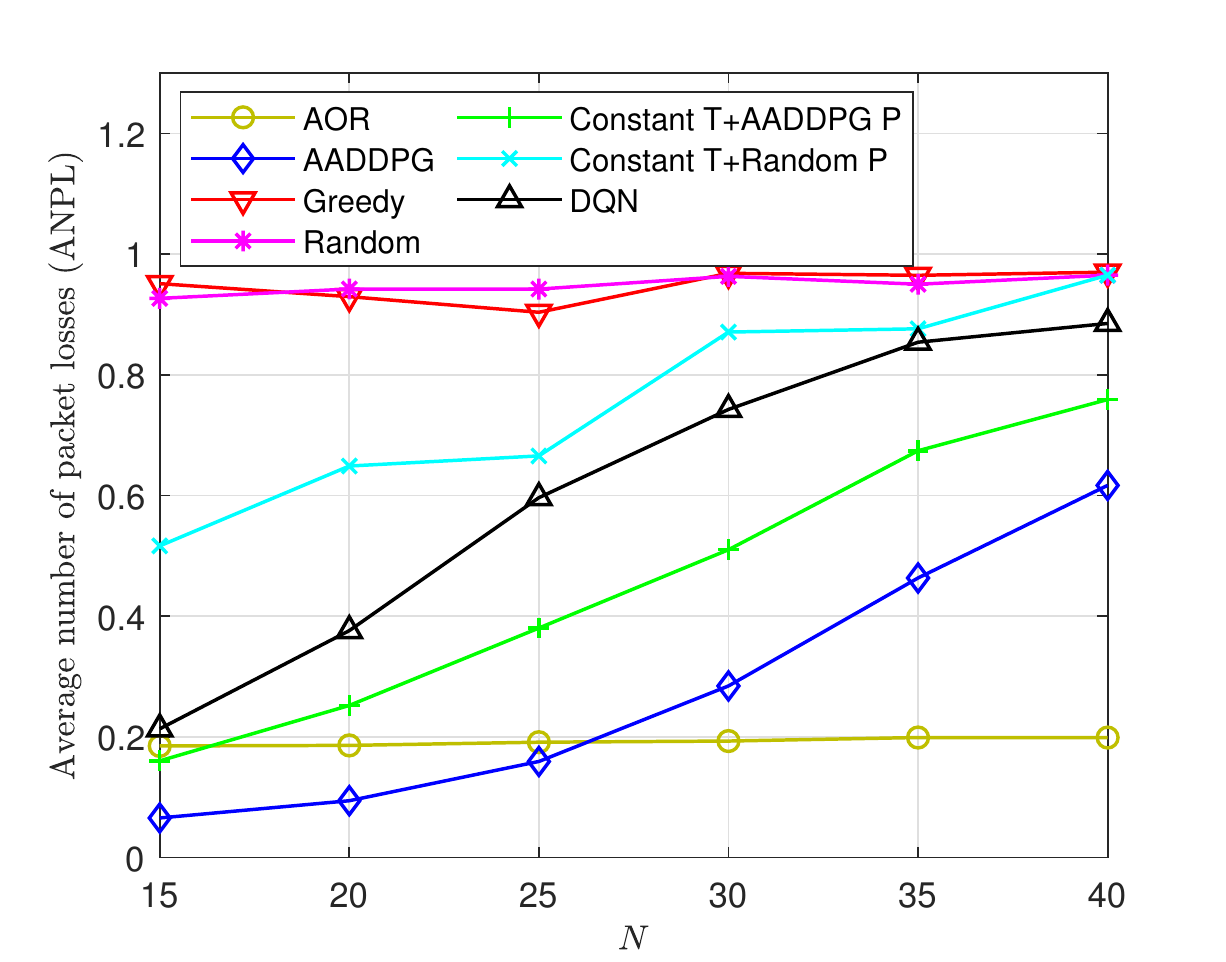}
	\caption{Impact of $N$ on the performance of the proposed AADDPG algorithm.}
	\centering
	\label{n}	
\end{figure}

Fig. \ref{lam} - Fig.\ref{n}  show the performance comparison between AADDPG algorithm and other benchmark algorithms with the change of system parameters. 
Fig. \ref{lam} plots the performance of   ANPL  versus data arrival rate $\lambda$ with $M=3$, $N=20$ and $B_{max}=6$. It can be observed that the performance of all schemes is degraded by increasing $\lambda$. This is because the larger $\lambda$ is, the more packets may arrive at each time slot and the more packets may be lost. The  random and greedy strategies have more than 90\% of the average number of loss packet with all $\lambda$. In addition, the  ANPL performance of AADDPG algorithm, constant $T$ + AADDPG $P$ algorithm and constant $T$ + Random $P$ algorithm almost increases linearly with the increase of $\lambda$, but obviously the slope of AADDPG algorithm is smaller, i.e., AADDPG has more obvious advantages with  larger $\lambda$.  In addition, it can be seen that the AADDPG algorithm performs much better than the DQN algorithm, especially when $\lambda > 1$. When $\lambda<0.8$, the performance of AOR is better than that of AADDPG, while with the increase of $\lambda$, its performance deteriorates more significantly and the performance gap with AADDPG increases gradually. This is because AOR sends data with the threshold $R^0$. When $\lambda$ increases, the packets cannot be transmitted in time, which causes a large number of packets losses . When $\lambda=2$, the  ANPL  performance of  constant $T$ + AADDPG $P$ algorithm and constant $T$ + random $P$ algorithm is 60\% and 130\% worse than AADDPG algorithm, respectively.

Fig. \ref{bmax} illustrates the impact of the buffer capacity size on different resource allocation algorithms with $\lambda=1$, $M=3$ and $N=20$. As shown in Fig., the  ANPL  performance of all schemes improves as $B_{max}$ increases, because the larger buffer space means that more packets can be temporarily stored to prevent data loss. Also, we can find that the AADDPG algorithm achieves almost 5\% packet loss when the $B_{max}\geq 8$, and even if $B_{max}=3$, the ANPL of the AADDPG algorithm is less than 30\%.  And the performance of  the constant $T$ + AADDPG $P$, constant $T$ + Random $P$, and DQN algorithm is about 50\%, 160\%, 150\% worse than AADDPG, respectively. The performance of AOR is inferior to that of AADDPG. This is because when  $\lambda=1$, the AOR algorithm is unable to transmit all the arriving data in time. 

In Fig. \ref{n}, the impact of  $N$  is demonstrated with $M=3$, $\lambda=1$, and $B_{max}=6$. The random and greedy algorithms still have the worst performance, and the performance of AADDPG and the other two benchmark algorithms  gets worse as $N$ increases. This is attributed to the fact that more  SSUs cause more interference, which leads to a decrease in the rate of all users and further prevents more packets from being transmitted in time. However, we can be seen that compared to the  Constant $T$ + AADDPG $P$, Constant $T$ + Random $P$, and DQN algorithms, the performance of the AADDPG deteriorates more slowly with increasing $N$  and even when $N = 40$, it still yields an  ANPL of about $0.61$.  The performance of the constant $T$ + AADDPG $P$, constant $T$ + Random $P$, and DQN algorithm is about 23\%, 56\%, and 43\% worse than AADDPG, respectively. In addition, we find that the performance of AOR almost does not change as $N$ increases, and is better than that of  the AADDPG algorithm when $N>25$. This is because all accessed SSUs in the AOR are transmitting data at the lowest transmit power that can satisfy the  threshold $R^0$, which guarantees that the algorithm can access more SSUs at the same time. However,  the AOR algorithm is realized with the premise that the global information is known, which is unrealistic for dynamic EH wireless communication systems.

\subsection{Algorithm Performance Verification}
In this section, we will examine the performance of the AADDPG algorithm as multiple system parameters change simultaneously. This is important to the practical application of the proposed algorithm, and network operators can better tune the network parameters to achieve the best system performance based on the study in this section.

\begin{figure}[t]
		\centering
	\includegraphics[width=4.5in]{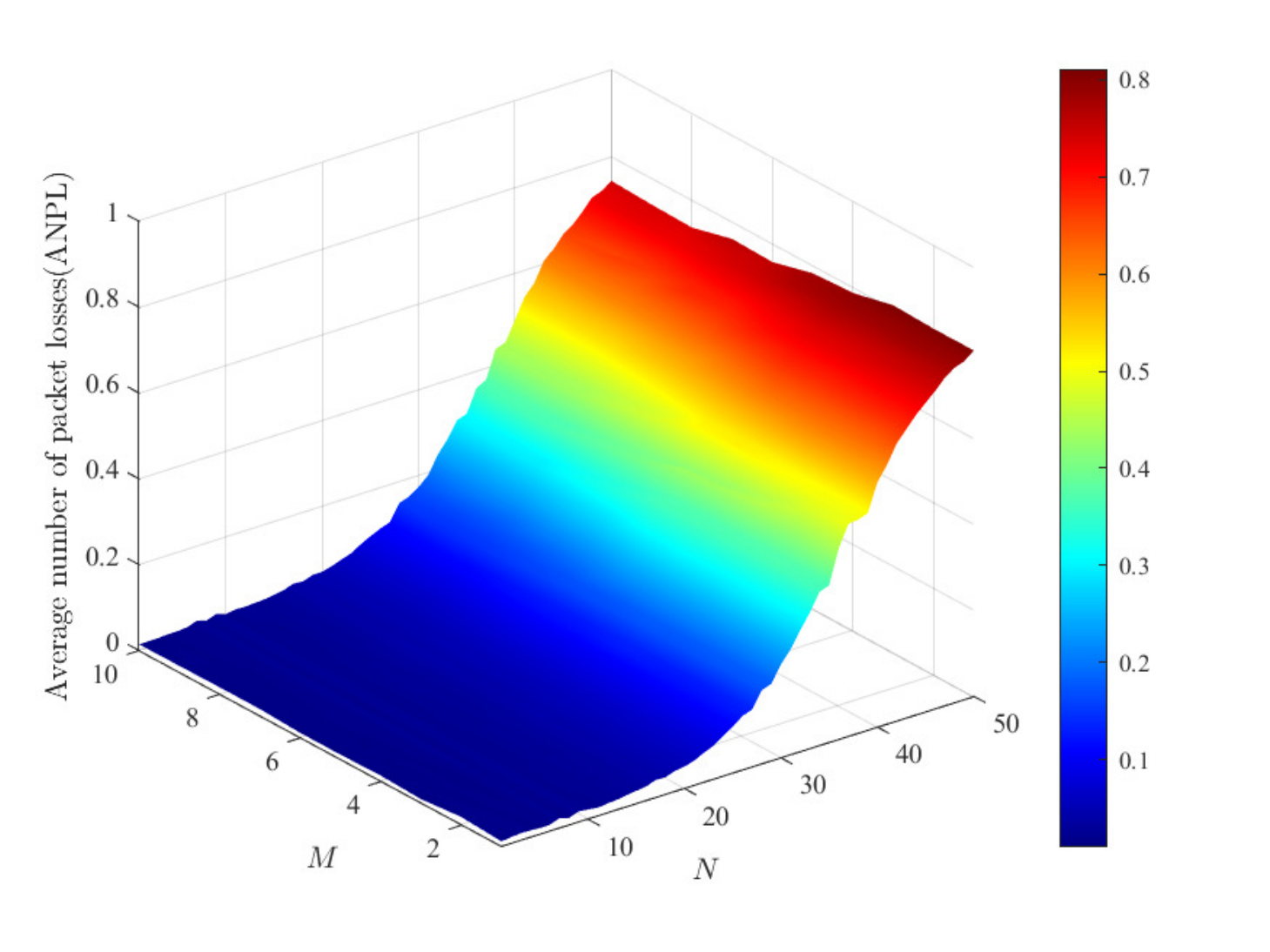}
	\caption{  ANPL under different number of SSUs and PUs.}
	\centering
	\label{NM}	
\end{figure}

Fig. \ref{NM} illustrates the  ANPL performance with different numbers
of SSUs and PUs with $B_{max}=8$, and $\lambda=1$. As shown in the figure, increasing the number of SSUs from $N = 1$ to $N = 50$, the  ANPL   performance of the system deteriorates, the value of  ANPL stay within 0.05 when the number of SSUs is within 20, and it exceed 0.65  when $N\geq45$. It is also observed that when the number of SSU is the same, there is an improvement in  ANPL  performance as the number of PUs $M$ increases, because more PUs means more energy sources. Also since the PUs are much closer to the base station compared to the SSUs, all the interfering signals from the PUs to the SSUs will be eliminated by the interference cancellation technique with a high probability.

\begin{figure}[t]
	\centering
	\includegraphics[width=4.5in]{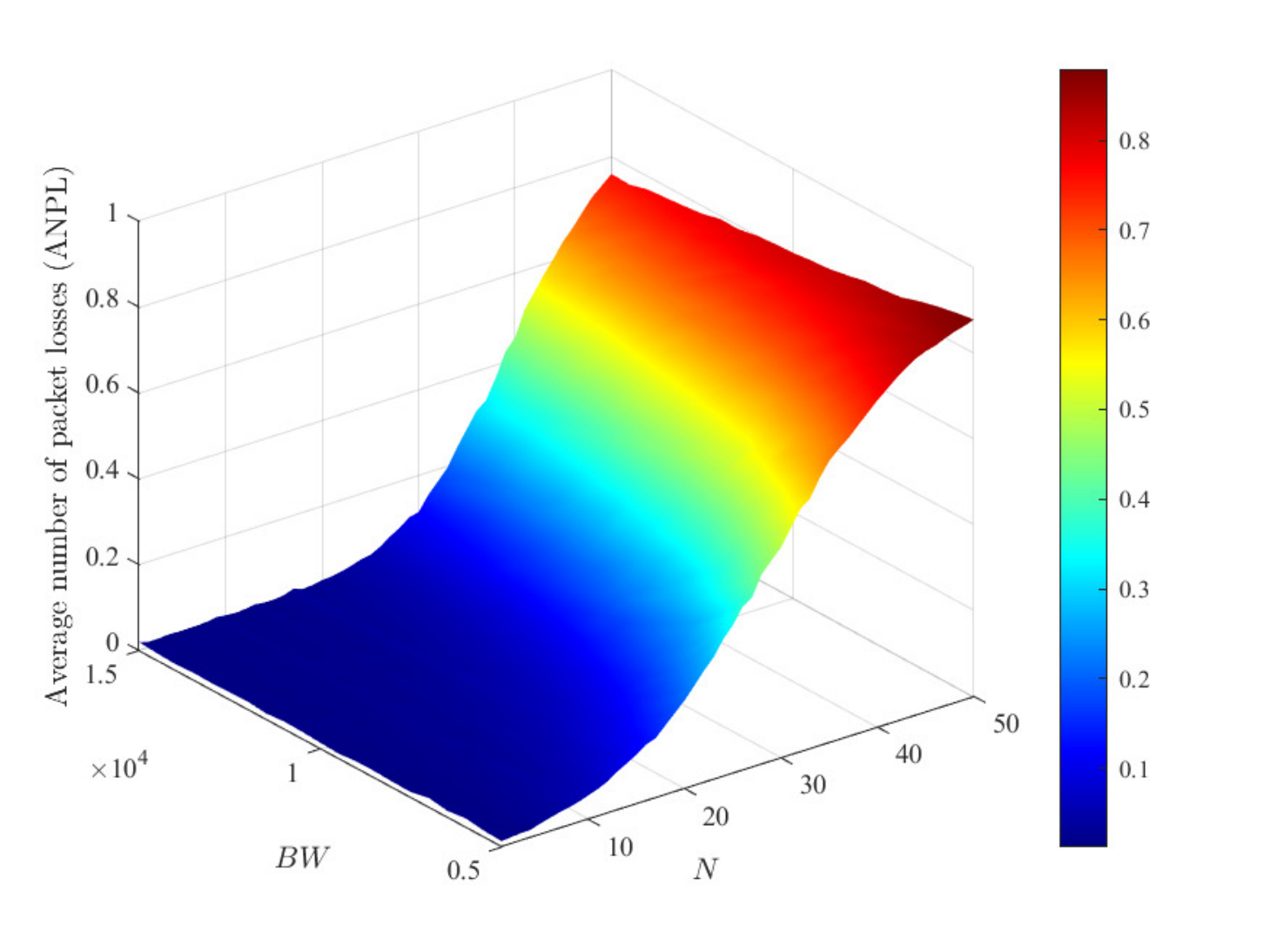}
	\caption{ ANPL under different number of SSUs $N$ and bandwidth $BW$.}
	\centering
	\label{Nbw}	
\end{figure}

Fig. \ref{Nbw} shows the relationship between  ANPL  and the number of SSUs N and the total bandwidth  of the system, where the number of SSUs $N$ increases from 1 to 50 and the total bandwidth BW increases  from $BW=5K$ Hz to $BW=15K$ Hz. It can be seen that with an increase of bandwidth, the average number of packet losses per user decreases gradually,  since more bandwidth leads to a higher data rate, and thus more packets can be transmitted in a timely manner. Based on the figure, the network operator can reasonably adjust the number of SSUs that can be accommodated in the system according to the current bandwidth and the  ANPL value that SSUs can  tolerated.  For example, if the bandwidth provided by the current system is $BW=12K$ Hz, the required data loss rate is ${ANPL}<0.2$, then the EH-CR-NOMA system can accommodate up to 31 SSUs.

\begin{figure}[t]
	\centering
	\includegraphics[width=4.5in]{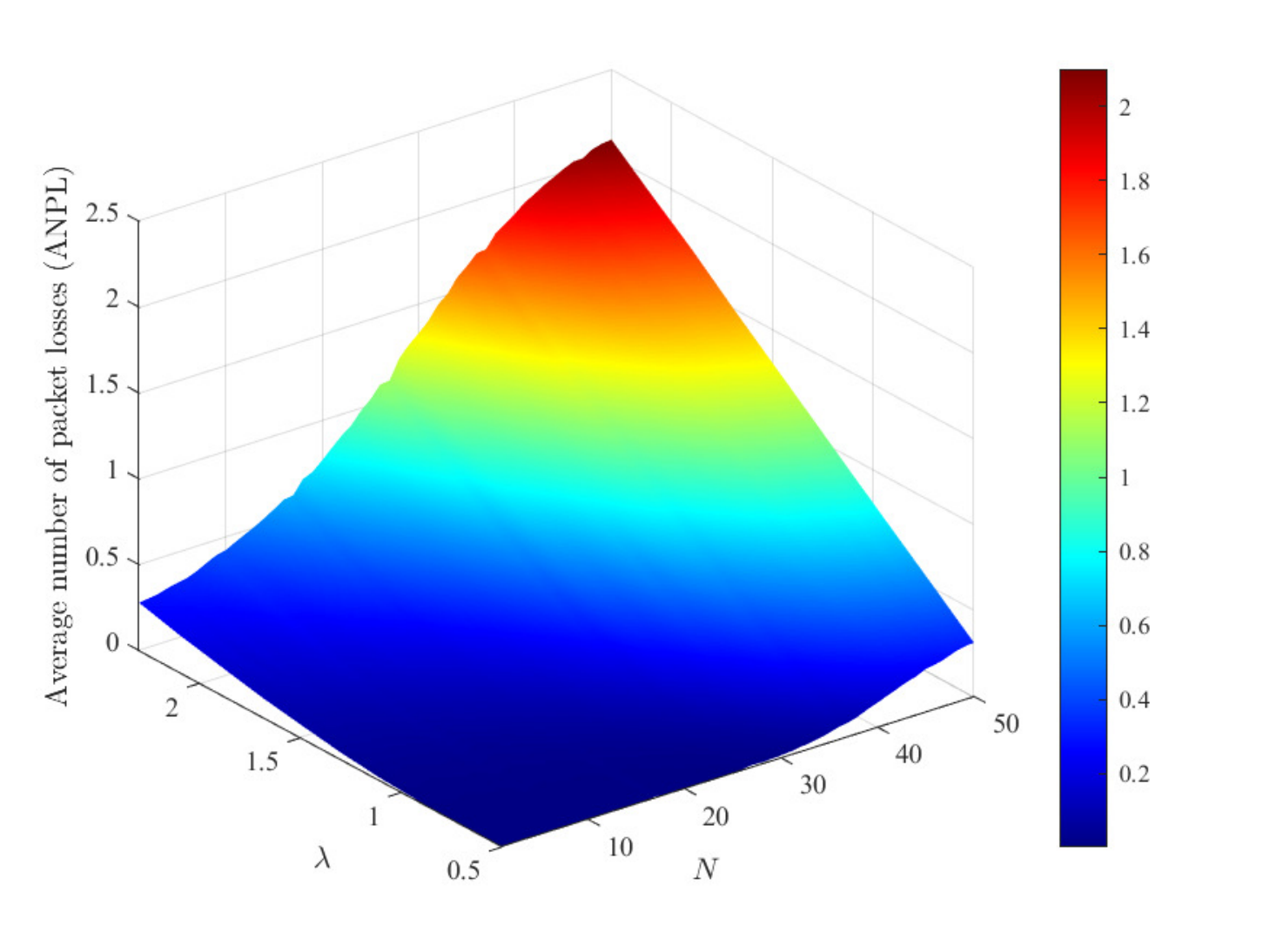}
	\caption{  ANPL under different number of SSUs and $\lambda$.}
	\centering
	\label{Nlam}	
\end{figure}

In Fig. \ref{Nlam} depicts the impact of the number of SSUs $N$ and packet arrival rate $\lambda$ on system ANPL with $B_{max}=8$, $M=3$, and $BW=10K$ Hz. Increasing the $\lambda$ from $0.5$ to $2.5$, as expected, for the same arrival rate $\lambda$,  the larger the number of SSUs in the system, the worse the ANPL performance, and for the same $N$,  ANPL deteriorates  as $\lambda$ increases, when $\lambda=0.5$, even if the number of SSUs $N=50$, the ANPL of 0.3 can still be obtained, while for $\lambda=2.5$, about 2 packets will be lost and lost for each SSU in each time slot.  Therefore, for the practical application of the proposed AADDPG algorithm, to prevent a large number of packets losses, the network operator needs to adjust the number of SSUs in time according to the arrival rate of packets in the system.

\begin{figure}[t]
	\centering
	\includegraphics[width=4.5in]{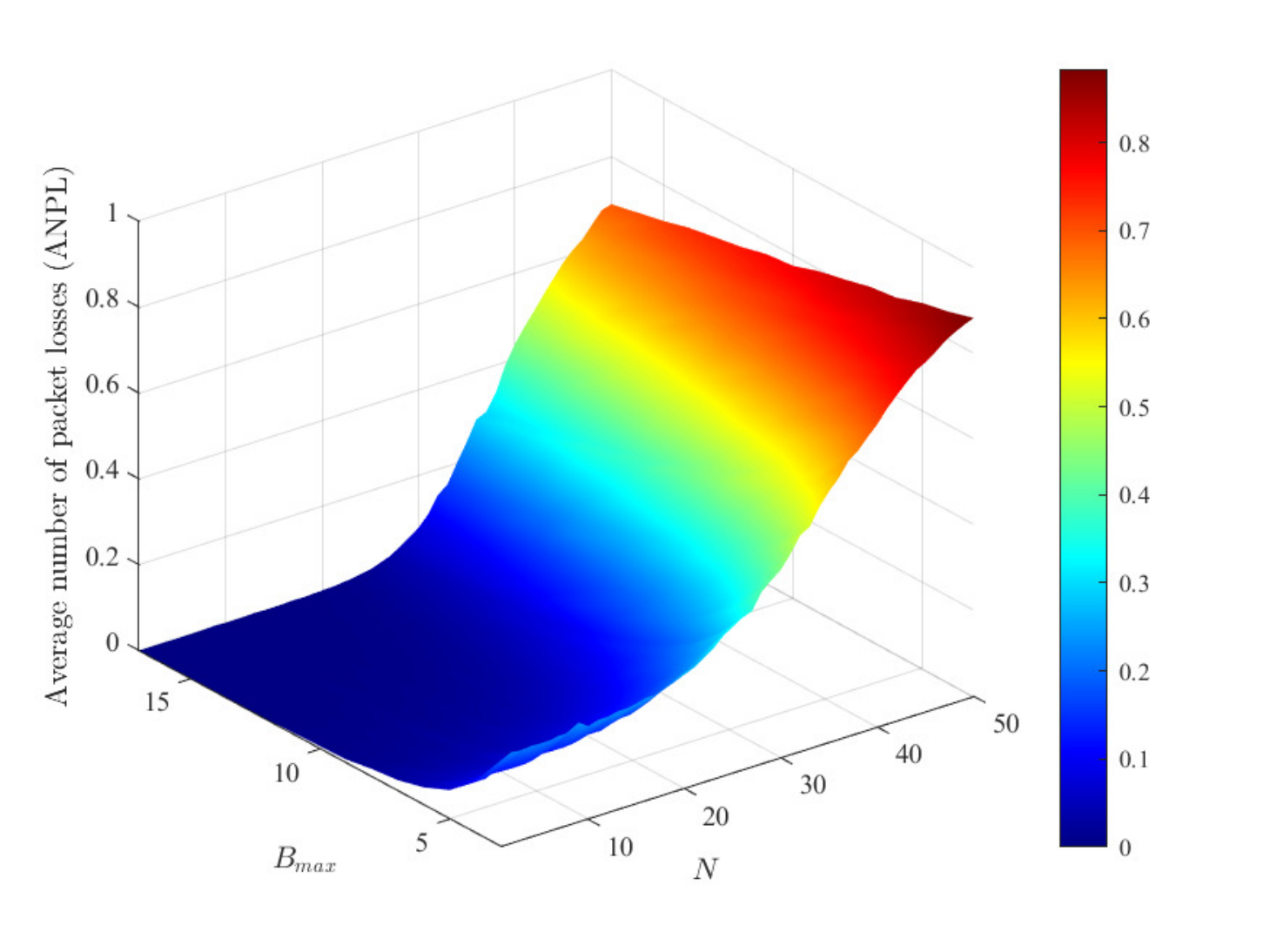}
	\caption{  ANPL under different number of SSUs and $\lambda$.}
	\centering
	\label{bmax1}	
\end{figure}

In Fig. \ref{bmax1}, we studied the effect of  $N$ and the buffer capacity $B_{max}$ of each SSU on ANPL with $M=3$, $\lambda=1$ and $BW=10K$ Hz. By increasing $B_{max}$ from $ 3$ to $ 17$, the  ANPL continuously improves, With $B_{max}\geq6$, each user will get an ANPL of less than 0.1, as long as the number of SSUs in the system does not exceed 20. However, as expected, the average number of  packets losses increases significantly as $N$ increases, or as $B_{max}$ decreases. Therefore, when the system must accommodate more users, a larger capacity buffer must be configured for each SSU in order to obtain better ANPL performance, e.g., when $N = 30$ and ANPL is required to be less than 0.1, the buffer capacity for each SSU must satisfy $B_{max}\geq15$.

\section{Conclusions}
In this paper, we have investigated the multidimensional joint resource scheduling problem for the EH- CR-NOMA system. In order to minimize the number of packets losses of SSUs under the constraints of energy, transmit power, and QoS requirements of both types of users, we designed a DDPG-based resource management algorithm to schedule time resources (time sharing factor of communication and energy collection), collected energy resources (transmitted power) and frequency resources (dynamic spectrum access of users) simultaneously by designing reasonable state space, reward function and action space. In addition, an action adjuster was introduced to achieve further improvements  the convergence and performance of the algorithm. Finally, sufficient experiments were performed to verify that the proposed AADDPG is able to achieve much lower packet loss relative to other benchmark algorithms. In the future, the problem of multidimensional and dynamic resource management of EH-CR-NOMA systems with MIMO and  delay requirements will be considered to further improve the applicability of our work. In addition, we will conduct research on theoretical analysis of DRL-based wireless systems.
 \appendices

\end{document}